\documentclass[10pt]{article}

\usepackage{amsmath}

\usepackage{array}

\usepackage{appendix}

\usepackage{tocloft}                   

\usepackage{graphicx}

\usepackage{amsfonts}

\usepackage{amssymb}

\usepackage{mathrsfs}

\usepackage{yfonts}

\usepackage{euscript}

\usepackage{centernot}                 

\usepackage{ifsym}                     

\usepackage{upgreek}

\usepackage{mathtools}

\usepackage{color}

\usepackage{slantsc}
\usepackage{calligra}

\usepackage{bbold}          

\usepackage[T1]{fontenc}

\usepackage{epsf}

\usepackage{latexsym}

\usepackage{tipa}

\usepackage{makeidx}

\makeindex

\def\be{\begin{equation}}

\def\ee{\end{equation}}

\def\beq{\begin{equation}}

\def\eeq{\end{equation}}

\def\bea{\begin{eqnarray}}

\def\eea{\end{eqnarray}}

\def\ni{\noindent}

\def\!{\hspace{-1.6667em}}

\def\mC{\mbox{C}}                        

\def\mM{\mbox{M}}                        

\def\mm{\mbox{m}}

\def\ms{\mbox{s}}

\def\uR{\underline{\mbox{$R$}}}

\def\uq{\underline{\mbox{q}}} 

\def\ur{\underline{\mbox{r}}}

\def\ux{\underline{\mbox{x}}}

\def\urho{{\underline{\rho}}}

\def\sm{\mbox{\scriptsize m}}

\def\so{\mbox{\scriptsize o}}

\def\sss{\mbox{\scriptsize s}}  

\def\sC{\mbox{\scriptsize C}}

\def\sI{\mbox{\scriptsize I}}

\def\sM{\mbox{\scriptsize M}}

\def\sS{\mbox{\scriptsize S}}

\def\sT{\mbox{\scriptsize T}}

\def\bigupalpha{\mbox{\Large$\alpha$}}

\def\cr{\mbox{\scriptsize{\bf $\mbox{ } \times \mbox{ }$}}}

\def\sumi2{\sum\mbox{}_{\mbox{}_{\mbox{\scriptsize $i$=1}}}^2}

\def\sumi3{\sum\mbox{}_{\mbox{}_{\mbox{\scriptsize $i$=1}}}^3}

\def\sumABcycles3{\sum\mbox{}_{\mbox{}_{\mbox{\scriptsize cycles $A,B$=1}}}^{3}}

\def\sumCDcycles3{\sum\mbox{}_{\mbox{}_{\mbox{\scriptsize cycles $C,D$=1}}}^{3}}

\def\sumj3{\sum\mbox{}_{\mbox{}_{\mbox{\scriptsize $j$=1}}}^3}

\def\sumk3{\sum\mbox{}_{\mbox{}_{\mbox{\scriptsize $k$=1}}}^3}

\def\prodiA1{\prod\mbox{}_{\mbox{}_{\mbox{\scriptsize $i$=1}}}^{A - 1}}

\def\Hilb{\mbox{{\boldmath$\mathfrak{H}$}ilb}}                 

\def\Phase{\mbox{{\boldmath$\mathfrak{P}$}hase}}                     

\def\bFrR{\mbox{\boldmath$\mathfrak{R}$}}                            
\def\Rig-Phase{\bFrR\mbox{ig-}\Phase}                                

\def\Positive-Modespace{\mbox{{\boldmath$\mathfrak{M}$}odespace$^+$}}

\def\POSITIVE-MODESPACE{\mbox{{\boldmath$\mathfrak{M}$}ODESPACE$^+$}}

\def\Kin-Hilb{\mbox{{\boldmath$\mathfrak{K}$}in-\Hilb}}                     

\def\Mid-Hilb{\mbox{{\boldmath$\mathfrak{M}$}id-\Hilb}}                     

\def\Dyn-Hilb{\mbox{{\boldmath$\mathfrak{D}$}yn-\Hilb}}                     

\def\5Star{\mbox{\Large$\star$}}              

\begin{document}

\begin{center}

{\Large \bf Two new versions of Heron's Formula}

\vspace{0.1in}

{\large \bf Edward Anderson$^*$}

\vspace{.2in}

\end{center}

\begin{abstract}

\ni Recollect that Heron's formula for the area of a triangle given its sides has a counterpart given the medians instead, which carries an extra factor 
of $\frac{4}{3}$.   
On the one hand, we formulate the pair of these in Linear Algebra terms, showing that they are related by a sides-to-medians involution $J$, 
which we find to furthermore commute with the `Heron map' $H$ as visible in the expanded version of Heron. 
Upon further casting the pair of these in terms of mass-weighted Jacobi coordinates, 
we find moreover that they are placed on an exactly equal footing, the factor of $\frac{4}{3}$ having now cancelled out. 
This motivates the `Heron--Jacobi' version of Heron's formulae, for mass-weighted area in terms of mass-weighted sides and mass-weighted 
medians respectively.  

\mbox{ }

\ni On the other hand, we show that diagonalizing the Heron map $H$ provides new derivations of, firstly, the famous Hopf map, 
                                              and, secondly, of Kendall's Theorem that the space of triangles is a sphere.  
This occurs by the `Heron--Hopf' version of Heron's formula simplifying down to none other than the on-sphere condition. 
Thus we establish that -- both an important fibre bundle model, and a foundational theorem of Shape Theory: a widely applicable Differential Geometry and Topology topic --   
arise together as consequences of just Heron's formula, Stewart's Theorem, and some elementary Linear Algebra manipulations.  
This working also accounts for the extra factor of 4 in the Hopf coordinate that is elsewise equal to the mass-weighted area in the 3-body problem context.  
It finally offers a new interpretation of the shape-theoretic ellipticity and anisoscelesness which realize the other two Hopf quantities:  
as eigenvectors shared by the Heron map $H$ and the sides-to-medians involution $J$.  

\end{abstract}

\mbox{ }

\ni Mathematics keywords: Applied Geometry, triangles, spaces of triangles, Shape Geometry, Shape Statistics, relative Jacobi coordinates, Hopf fibration.

\vspace{0.1in}
  
\ni $^*$ Dr.E.Anderson.Maths.Physics@protonmail.com

\section{Introduction}
%
{            \begin{figure}[!ht]
\centering
\includegraphics[width=0.17\textwidth]{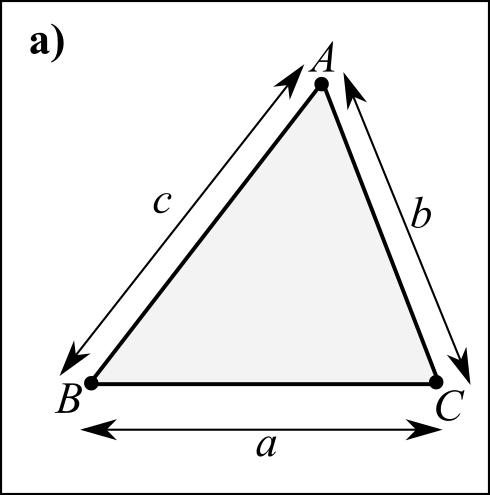}
\caption[Text der im Bilderverzeichnis auftaucht]{        \footnotesize{a) Labelling of vertices and edges of the triangle.  }  } 
\label{Triangle-Notation} \end{figure}           }

\ni{\bf Definition 1} Consider an arbitrary triangle $\triangle\,ABC$, denoted as in Fig \ref{Triangle-Notation}.   
Using $a_i$, $i = 1$ to $3$ to denote $a$, $b$, $c$ will also be useful for us.   

\mbox{ } 

\ni{\bf Definition 2} The {\it perimeter} $P$ is 
\be
P := a + b + c \mbox{ } = \mbox{ } \sum_i a_i \mbox{ } ,  
\ee
and the {\it semi-perimeter} is  
\be
\ms \mbox{ } := \mbox{ } \frac{P}{2} \mbox{ } = \mbox{ } \frac{a + b + c}{2} \mbox{ } = \mbox{ } \frac{1}{2}\sum_i a_i \mbox{ } .  
\ee
\ni{\bf Definition 3} We use $\mbox{Area}(ABC)$, or $\mbox{Area}$ for short when unambiguous, to denote the area of $\triangle\,ABC$.

\mbox{ }

\ni{\bf Theorem 1 (Heron's formula)}  
\be
Area = \sqrt{s(s - a)(s - b)(s - c)} \mbox{ } . 
\label{Heron}
\ee 
This is a classical result, known since the first century A.D. \cite{Hero}; see e.g.\ \cite{Coxeter} for a modern-era proof.   

\mbox{ }

\ni{\bf Corollary 1 (Expanded Heron's formula)}  
\be
Area \mbox{ } =  \mbox{ }\frac{1}{4}\sqrt{(2 \, a^2 b^2 - c^4) + cycles} \mbox{ } . 
\label{Exp-Heron} 
\ee
\ni{\bf Outline of the rest of this Paper}

\mbox{ }

\ni In Sec 2, we re-express (\ref{Exp-Heron}) in Linear Algebra terms, involving what we term the `Heron matrix', $\underline{\underline{H}}$.  
In Sec 3, we recollect that side lengths control median lengths and vice versa, 
via a corollary \cite{IMO} of {\it Stewart's Theorem} \cite{Stewart, PS70} (another classical result, now from the 18th century).
We recast this inter-relation also in Linear Algebra terms, 
showing moreover that it can be formulated as an involution $\underline{\underline{J}}$: the {\it sides-medians involution}.    
Perhaps surprisingly, $\underline{\underline{H}}$ and $\underline{\underline{J}}$ are furthermore shown to commute.
This accounts for why the usual side's Heron formula and the medians' Heron formula (\cite{Benyi}, Sec \ref{Med-Heron}) are very similar in appearance, 
differing only by a relative factor of $\frac{4}{3}$. 

\mbox{ }

\ni We introduce 3-particle relative Jacobi coordinates in Secs 4 and 5.  
These are well-known to be useful in the $N$-body problem context \cite{Marchal, LR97}.
For the equal point masses case currently under consideration, medians enter Jacobi coordinates on the same footing as sides.
This gives the first reason -- Jacobian motivation -- why I am reappraising the theory of medians, and,  
more specifically, am considering the theory of Jacobi mass-weighted medians on the same footing as mass-weighted sides.
We show moreover in Sec 6 that the Jacobi mass-weighted side and median forms of Heron's formulae -- which we term {\it `Heron--Jacobi' formulae} -- 
are now on an {\it identical} footing to each other, the factor of $\frac{4}{3}$ having been absorbed and given a new conceptual identity in the process.

\mbox{ }

\ni In Sec 7, we furthermore consider diagonalizing the Heron matrix $\underline{\underline{H}}$.   
We observe this to give none other than a recovery of the famous {\it Hopf map} \cite{Hopf} (Appendix A), 
which, in the present 2-$d$ 3-body problem context \cite{Dragt, Iwai87, LR97, FileR, I, III} (Appendix B), 
is also a way of obtaining {\it Kendall's Theorem} \cite{Kendall84, Kendall89, Kendall} that the space of all triangular shapes is a sphere.
This is via the {\it `Heron--Hopf'} version of Heron's formula having reduced to what is mathematically just the on-sphere condition. 
This approach moreover builds upon the preceding use of Relative Jacobi coordinates, 
which are thus useful in setting up David Kendall's Shape Theory: 
a new subject of considerable promise \cite{Kendall89, Small, Kendall, FileR, Bhatta, QuadI, AMech, PE16, ABook, MIT, I, II, III, IV, A-Pillow}.

\mbox{ }

\ni This working also accounts for the extra factor of 4 in the Hopf coordinate that, in the 3-body problem context, is mass-weighted area up to this factor.
It furthermore points to the other two Hopf coordinates 
-- interpreted in \cite{+Tri, FileR, III} in the 3-body problem context as {\it ellipticity} and {\it anisoscelesness} -- 
having comparable status to the much more well-known area variable. 
This is from the point of view of these two variables featuring co-primally alongside the area as a set of three Cartesian axes for the shape sphere's natural ambient $\mathbb{R}^3$. 
It additionally offers a new interpretation of the shape-theoretic ellipticity and anisoscelesness realizations of Hopf's other two quantities.
Namely, these are none other than the Heron map   $\underline{\underline{H}}$'s eigenvectors, which, by commutativity, 
are shared also with the sides--median involution $\underline{\underline{J}}$.   
Following some further Linear Algebra consideration -- now of the Hopf quantities -- in Sec 8, 
we show in Sec 9 that the ellipticity and anisoscelesness quantities to be invariant in form under exchange of sides and medians, 
up to signs which are allowed as part of choosing Cartesian axes.
This provides our second -- now `Hopfian' -- motivation for treating sides and medians on an equal footing.
Sec 10 finally summarizes what we term the {\it `Heron--Hopf'} and {\it `Heron--Kendall--Hopf'} forms of Heron's formula -- the area-subject and symmetrical presentations --
alongside giving two (almost) equivalent concomitant formula with ellipticity and anisoscelesness as their subjects respectively.  

\mbox{ }

\ni As complementary reading, see \cite{Kendall, I, III} for accounts of Kendall's Shape Theory in general and shape space of triangles in particular, 
and also \cite{III} for an outline of the Hopf map and its realization in the Shape Theory of triangles.

\section{The Heron matrix}

\ni{\bf Lemma 1} The expanded form (\ref{Exp-Heron}) of Heron's formula (\ref{Heron}) can be recast in Linear Algebra terms as the quadratic form 
-- in squares $a_i\mbox{}^2$, so it is quartic in the $a_i$ themselves --
\be 
(4 \times \mbox{Area})^2 = H_{ij} a_i\mbox{}^2 a_j\mbox{}^2
\label{Heron-form}
\ee 
for `{\it Heron matrix}' 
\be
\underline{\underline{H}} := 
\frac{1}{3}   
\left( \stackrel{  \stackrel{  \mbox{$   -              1  \mbox{ } \mbox{ } \mbox{ }  1 \mbox{ } \mbox{ } \mbox{ }  1$}  }  
                                         {  \mbox{$\mbox{ } \mbox{ } \, 1  \mbox{ }    -               1 \mbox{ } \mbox{ } \mbox{ }  1$}  }  }
										 {  \mbox{$\mbox{ } \mbox{ } \, 1  \mbox{ } \mbox{ } \mbox{ }  1 \mbox{ }    -               1$}  } \right)  \mbox{ } .
\ee
\ni{\bf Remark 1} A more conceptual name for the formula habitually named after Heron is `{\it area from side data} formula'.

\section{The sides--medians involution}\label{Med-Heron}

\ni{\bf Definition 4} The {\it medians} of a triangle areas per \ref{Medians}.a)-b).  
It will also be useful for us to use $m_i$, $i = 1$ to $3$ to denote $m_a$, $m_b$, $m_c$.  
%
{            \begin{figure}[!ht]
\centering
\includegraphics[width=0.51\textwidth]{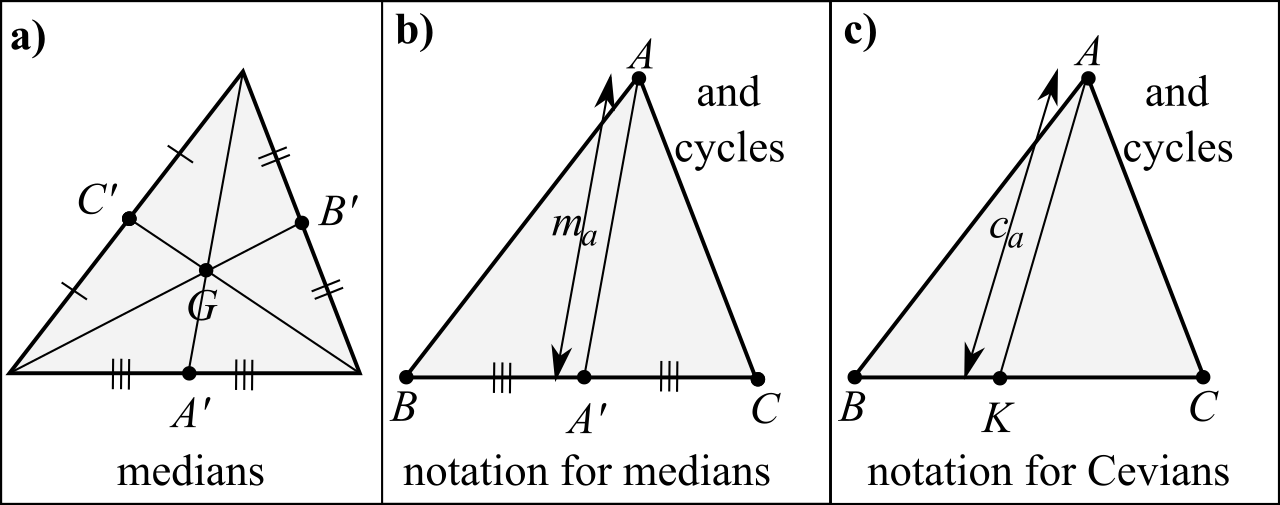}
\caption[Text der im Bilderverzeichnis auftaucht]{        \footnotesize{a) Definition of the medians, concurring at the {\it centroid}, alias {\it centre of mass}, $G$.
b) Further notation for the medians. 
c) General Cevian set-up, of which medians are a particular subcase.}  } 
\label{Medians} \end{figure}           }

\mbox{ }

\ni{\bf Remark 2} By treating the sides and the medians on an equal footing, 
we have more definitions (or at least accordances of equal significance) than in hitherto standard treatments of triangles, starting with the following.  

\mbox{ }

\ni{\bf Definition 5} The {\it medimeter} $M$ is 
\be
M := m_a + m_b + m_c \mbox{ } = \mbox{ } \sum_i m_i \mbox{ } ,   
\ee
whereas the {\it semi-medimeter} is   
\be
s = \frac{M}{2} \mbox{ } .  
\ee
\ni{\bf Remark 3} The perimeter and medimeter can furthermore be viewed as first moments of sides and medians respectively.
The second moments counterparts of each of these also enter the current paper, as follows. 

\mbox{ }

\ni{\bf Definition 6} The {\it second moment of sides} is 
\be
P_{\sI\sI} := a^2 + b^2 + c^2 \mbox{ } = \mbox{ } \sum_i a_i\mbox{}^2 \mbox{ } ,   
\ee
and the {\it second moment of medians} is 
\be
M_{\sI\sI} := m_a\mbox{}^2 + m_b\mbox{}^2 + m_c\mbox{}^2 \mbox{ } = \mbox{ } \sum_i m_i\mbox{}^2 \mbox{ } .    
\label{MII}
\ee
\ni{\bf Theorem 2 (Stewart's Theorem)} Let $\triangle ABC$ be a triangle with $K$ an arbitrary point on side $BC$.  
Then
\be
{AK}^2 \mbox{ } = \mbox{ } \frac{KC}{BC} \, {AB}^2 \mbox{ } + \mbox{ } \frac{BK}{BC} \, {AC}^2 - BK \, KC \mbox{ } .  
\ee 
\ni{\bf Proof} See \cite{PS70}. $\Box$ 

\mbox{ }

\ni{\bf Remark 4} The segment $AK = c_a$ is in general defined to be a {\it Cevian} (Fig \ref{Medians}.c) -- a basic building block for affine-geometric considerations -- 
as enter Ceva's own theorem \cite{PS70, Silvester}.  
In Euclidean geometry, moreover, Cevians have lengths as well as affine properties, and the function of Stewart's Theorem is to compute the former. 
Thus a more conceptual name for Stewart's Theorem is {\it Cevian length Theorem}.   

\mbox{ }

\ni{\bf Remark 5} Medians are indeed a simple subcase of Cevians, hence the relevance of Stewart's Theorem to the current paper, as follows.  

\mbox{ }

\ni{\bf Corollary 2} i) The median lengths' squares are given by 
\be
m_a\mbox{}^2 \mbox{ } = \mbox{ } \frac{  2 \, b^2 + 2 \, c^2 - a^2  }{  4  } \mbox{ } \mbox{ and cycles } .
\label{0.1}
\ee
\ni ii) The second moments of medians and of sides are related by 
\be 
M_{\sI\sI} \mbox{ } = \mbox{ } \frac{3}{4} \, P_{\sI\sI} \mbox{ } .
\label{0.2}
\ee
\ni{\underline{Proof}} i)  This readily follows from Stewart's Theorem, as per worked problem 1 of \cite{IMO}.   

\mbox{ }

\ni ii) then follows immediately from both parts of Definition 6 upon summing i) over all cycles. $\Box$

\mbox{ }

\ni{\bf Corollary 3} i) In Linear Algebra form, 
\be
             \left( \stackrel{  \stackrel{  \mbox{$m_a\mbox{}^2$}  }{  \mbox{$m_b\mbox{}^2$}  }  }{\mbox{$m_c\mbox{}^2$}} \right) \mbox{ } = \mbox{ } 
\frac{1}{4}  \left( \stackrel{  \stackrel{  \mbox{$   -              1  \mbox{ } \mbox{ } \mbox{ }  2 \mbox{ } \mbox{ } \mbox{ } \mbox{ }  2$}  }  
                                         {  \mbox{$\mbox{ } \mbox{ } \mbox{ } 2  \mbox{ }    -               1 \mbox{ } \mbox{ } \mbox{ }  2$}  }  }
										 {  \mbox{$\mbox{ } \mbox{ } \, 2  \mbox{ } \mbox{ } \mbox{ }  \mbox{ } 2 \mbox{ }    -               1$}  } \right) 
										 \mbox{ } = \mbox{ } 
             \left( \stackrel{\stackrel{\mbox{$a^2$}}{\mbox{$b^2$}}}{\mbox{$c^2$}} \right) \mbox{ } , 
\label{0.3}
\ee
i.e.\ 
\be
m_i\mbox{}^2 \mbox{ } = \mbox{ } \frac{1}{4} \, B_{ij} a_j\mbox{}^2 \mbox{ } ,
\label{0.4}
\ee
for  
\be
\underline{\underline{B}} \mbox{ } := \mbox{ } 
\left( \stackrel{  \stackrel{  \mbox{$   -              1  \mbox{ } \mbox{ } \mbox{ }  2 \mbox{ } \mbox{ } \mbox{ } \mbox{ }  2$}  }  
                                         {  \mbox{$\mbox{ } \mbox{ } \mbox{ } 2  \mbox{ }    -               1 \mbox{ } \mbox{ } \mbox{ }  2$}  }  }
										 {  \mbox{$\mbox{ } \mbox{ } \mbox{ } 2  \mbox{ } \mbox{ } \mbox{ }  \mbox{ } 2 \mbox{ }    -               1$}  } \right)
\label{0.5}
\ee
a symmetric matrix 
\be
\underline{\underline{B}} = \underline{\underline{B}}^{\sT} \mbox{ } , \mbox{ } \mbox{ i.e.\ in components } \mbox{ } B_{ij} = B_{ji} \mbox{ } .
\label{Sym-B}
\ee
\ni ii) Inverting, 
\be
a_i\mbox{}^2 \mbox{ } = \mbox{ } \frac{4}{9} \, B_{ij} m_j\mbox{}^2 \mbox{ } . 
\label{0.6}
\ee
\ni{\bf Remark 6} That the same $B_{ij}$ appears in the inverted expression indicates that $B_{ij}$ is proportional to an {\it involution} $J_{ij}$, 
i.e.\ it is a matrix such that 
\be
\underline{\underline{J}}^2 = \underline{\underline{1}}: \mbox{ the identity matrix } .  
\label{0.7}
\ee
Thereby, we can further tidy up Corollary 1's Linear Algebra formulation by identifying and using $\underline{\underline{J}}$, as follows. 

\mbox{ }

\ni{\bf Corollary 4} i) 
\be
m_i\mbox{}^2 \mbox{ } = \mbox{ } \frac{3}{4} \, J_{ij} a_j\mbox{}^2 \mbox{ } , \mbox{ } \mbox{ and } 
\label{0.8}
\ee
ii) 
\be
a_i\mbox{}^2 \mbox{ } = \mbox{ } \frac{4}{3} \, J_{ij} m_j\mbox{}^2 \mbox{ } ,  
\label{0.9}
\ee
for {\it sides-medians involution}   
\be
\underline{\underline{J}} \mbox{ } := \mbox{ }  
\frac{1}{3}   
\left( \stackrel{  \stackrel{  \mbox{$   -              1  \mbox{ } \mbox{ } \mbox{ }  2 \mbox{ } \mbox{ } \mbox{ } \mbox{ }   2$}  }  
                                         {  \mbox{$\mbox{ } \mbox{ } \, 2  \mbox{ }    -               1 \mbox{ } \mbox{ } \mbox{ }  2$}  }  }
										 {  \mbox{$\mbox{ } \mbox{ } \, 2  \mbox{ } \mbox{ } \mbox{ }  2 \mbox{ }    -               1$}  } \right)  
										 \mbox{ } = \mbox{ } \frac{1}{3} \, \underline{\underline{B}} \mbox{ } ,
\ee
which of course remains symmetric, 
\be
\underline{\underline{J}} = \underline{\underline{J}}^{\sT} \mbox{ } , \mbox{ } \mbox{ i.e.\ in components } \mbox{ } J_{ij} = J_{ji} \mbox{ } .
\label{Sym}
\ee
\ni{\bf Theorem 3 (Medians' Heron formula)}, alias `area from median data' formula.
\be
\mbox{Area} \mbox{ } = \mbox{ } \frac{4}{3} \, \sqrt{\ms(\ms - m_a)(\ms - m_b)(\ms - m_c)} 
            \mbox{ } = \mbox{ } \frac{1}{3} \sqrt{(2 \, m_a\mbox{}^2 m_b\mbox{}^2 - m_c\mbox{}^4) + cycles} \mbox{ } . 
\label{Median-Heron}
\ee
{\underline{Proof}} While traditional geometric proofs of this are not uncommon \cite{Benyi}, I give instead a striking Linear Algebra proof.
First note Lemma 1's Linear Algebra form of the square of Corollary 1's expanded Heron formula.  
Next substitute Corollary 3.ii) in,  
\be
(4 \times \mbox{Area})^2 = a_i\mbox{}^2H_{ij}a_j\mbox{}^2
                         \mbox{ } = \mbox{ } \left( \frac{4}{3} J_{ik} m_k\mbox{}^2 \right)  H_{ij}  \left( \frac{4}{3} J_{jl} m_l\mbox{}^2 \right)
						 \mbox{ } = \mbox{ } \frac{16}{9} \, J_{ki} H_{ij} J_{jl} m_k\mbox{}^2 m_l\mbox{}^2 \mbox{ } ,
\ee 
where we used (\ref{Sym}) in the last step. 

\mbox{ }

\ni Thus evaluating the matrix product, 
\be 
\mbox{Area}^2 \mbox{ } = \mbox{ } \frac{1}{9} \, H_{ij} m_i\mbox{}^2 m_j\mbox{}^2 \mbox{ } , 
\label{LA-Median-Heron}
\ee 
so reversing the expansion of Heron with $m_i$ in place of $a_i$, (\ref{Median-Heron}) ensues. $\Box$ 

\mbox{ }

\ni{\bf Remark 7} This proof contains an insight which traditional geometric proofs miss. 
Namely, that the side--median involution matrix $\underline{\underline{J}}$ and the `Heron matrix' $\underline{\underline{H}}$ {\sl commute}, 
\be
\mbox{\bf [}\underline{\underline{J}} \mbox{\bf ,} \, \mbox{ } \underline{\underline{H}} \mbox{\bf ]} \mbox{ } = \mbox{ } 0 \mbox{ } , \mbox{ } \mbox{ i.e.\ } \mbox{ } 
\underline{\underline{J}} \, \underline{\underline{H}} = \underline{\underline{H}} \, \underline{\underline{J}} \mbox{ } .
\ee
Note moreover that 
\be
\underline{\underline{J}} \, \underline{\underline{H}} = \underline{\underline{K}} = \underline{\underline{H}} \, \underline{\underline{J}}  \mbox{ } \mbox{ for } \mbox{ }  \mbox{ }
\underline{\underline{K}} \mbox{ } := \mbox{ } \frac{1}{3} \, 
\left( \stackrel{  \stackrel{  \mbox{$   \mbox{ } \mbox{ }5  \mbox{ }   -1 \mbox{ }   -1$}  }  
                                         {  \mbox{$-1  \mbox{ } \mbox{ }  \mbox{ }                       5   \mbox{ }   -1$}  }  }
										 {  \mbox{$-1  \mbox{ }          -1   \mbox{ } \mbox{ } \mbox{ }                5$}  }     \right)     \mbox{ } . 
\label{K}
\ee
\ni{\bf Remark 8} It is because of this that the `median-Heron' matrix $\underline{\underline{M}}$ in the a priori conceptual form of (\ref{LA-Median-Heron}), 
\be 
Area^2 = M_{ij} m_i\mbox{}^2 m_j\mbox{}^2 \mbox{ } , 
\label{LA-Median-Heron-Conceptual}
\ee 
is just proportional to the `Heron matrix' itself, 
\be
\underline{\underline{M}} \mbox{ } = \mbox{ } \frac{1}{9} \, \underline{\underline{H}} \mbox{ } . 
\ee 
\ni{\bf Remark 9} In summary, the sides-Heron and medians-Heron formulae are 
\be
\sqrt{s(s - a)(s - b)(s - c)} = \mbox{Area} \mbox{ } = \mbox{ } \frac{4}{3} \sqrt{\ms(\ms - m_a)(\ms - m_b)(\ms - m_c)} \mbox{ } .  
\label{Summary-1}
\ee

\section{Jacobi coordinates for the triangle}
%
{            \begin{figure}[!ht]
\centering
\includegraphics[width=1.0\textwidth]{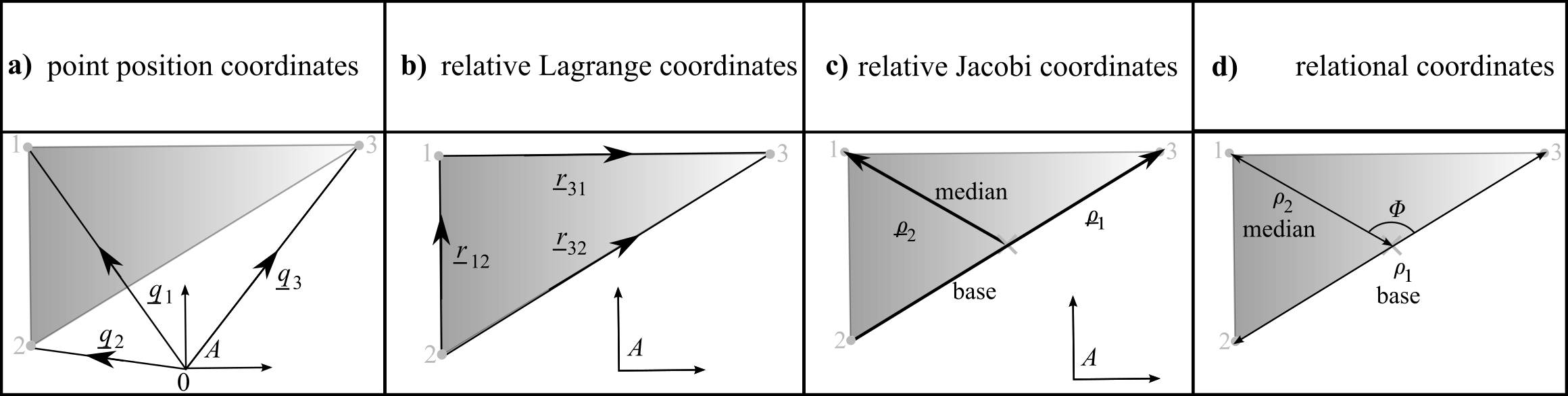}
\caption[Text der im Bilderverzeichnis auftaucht]{        \footnotesize{a) Point particle position coordinates.
b) Relative Lagrange separation vectors.   
c) Relative Jacobi separation vectors; the cross denotes the centre of mass of particles 2 and 3. 
d) Coordinates not depending on the absolute axes either: Jacobi magnitudes and the angle between them. 
To finally not depend on the scale, take the ratio $\rho_2/\rho_1$ of the Jacobi magnitudes alongside this angle. 
On the shape sphere, moreover, $\Phi$ plays the role of polar angle and the arctan of this ratio plays the role of azimuthal angle.} }
\label{Jac-Med-Ineq-Fig-2}\end{figure}            }

\ni{\bf Structure 1} Let us now consider our triangle's vertices to be equal-mass particles with position vectors $\uq_I$ ($I = 1$ to $3$) 
relative to an absolute origin 0 and axes $A$ (Fig \ref{Jac-Med-Ineq-Fig-2}.a).

\mbox{ }

\ni{\bf Definition 7} The {\it inertia quadric} for $N$ particles in any dimension is 
\be 
I(\underline{q}_I) \mbox{ } = \mbox{ } \sum_{I = 1}^N \mm_I q_I\mbox{}^2 
                   \mbox{ } = \mbox{ } \sum_{I = 1}^N     q_I\mbox{}^2 \mbox{ } .
\ee 
where the last equality is for equal masses, taken without loss of generality to be of unit size.

\mbox{ }

\ni{\bf Remark 10} Translating the origin by some arbitrary amount $\underline{a}$, 
\be
I(\underline{q}_I, \underline{a}) \mbox{ } = \mbox{ } \sum_{I = 1}^N \mm_I||\underline{q}_I - \underline{a}||^2 \mbox{ } .
\label{Ia}
\ee 
Furthermore, varying with respect to $\underline{a}$,   
\be
\underline{a} \mbox{ } = \mbox{ } \frac{1}{\mM} \sum_{I = 1}^N \mm_I q_I =: \underline{a}_{\sC\so\sM} \mbox{ }
\ee 
where `CoM' stands for `centre of mass', and where 
\be
\mM \mbox{ } := \mbox{ } \sum_{I = 1}^N \mm_I \mbox{ } 
\ee 
is the {\it total mass}.  
I.e.\ introducing an arbitrary $\underline{a}$ and varying picks out the centre of mass position. 
Then substituting for this back in (\ref{Ia}), one obtains the relative Lagrangian version of the inertia quadric, 
\be
I(\underline{r}_{IJ}) \mbox{ } = \mbox{ } \frac{1}{\mM}\stackrel{    \mbox{$\sum_{I = 1}^N\sum_{J = 1}^N$}    }{    \mbox{\scriptsize$I < J$}    } \, m_I m_J r_{IJ}\mbox{}^2 \mbox{ } ,
\label{ILag}
\ee 
for {\it relative Lagrange coordinates}
\be 
\underline{r}_{IJ} := \underline{q}_J - \underline{q}_I 
\ee 
(see Fig \ref{Jac-Med-Ineq-Fig-2}.b)).  
Formulating in terms of these, by virtue of their being differences of position vectors $\underline{q}_I$, cancels out any reference to the absolute origin, $0$.   

\mbox{ }

\ni{\bf Remark 11} For $N \geq 3$, this object has disadvantages stemming from not all the $\underline{r}_{IJ}$ being independent.
Non-diagonality ensues.
This can of course be circumvented by diagonalization, which, in this context, amounts to passing to {\it relative Jacobi coordinates}. 
These are moreover no longer in general inter-particle separations, but rather the broader concept of inter-particle {\sl cluster} separations. 
As we shall see below for the particular example of $N = 3$ in 2-$d$ -- the triangle -- this generalization involves relative separations between subsystem centres of mass. 
[This concept includes inter-particle separations by the identity that particle positions coincide with that 1-particle subsystem's centre of mass.]

\mbox{ }

\ni{\bf Remark 12} To proceed for our particular example, we rewrite (\ref{ILag}) in Linear Algebra form 
\be
I(\underline{r}_{IJ}) = L_{IJ} q_I q_J
\ee
for `relative Lagrange matrix' $\underline{\underline{L}}$.  
Narrowing down consideration to $N = 3$, 
\be
\frac{1}{3}(r_{12}\mbox{}^2 + r_{13}\mbox{}^2 +  r_{23}\mbox{}^2) \mbox{ } , 
\ee 
for which 
\be
\underline{\underline{L}} \mbox{ } = \mbox{ } 
\frac{1}{3} \left( \stackrel{  \stackrel{  \mbox{$\mbox{ } \mbox{ } 1  \mbox{ } \mbox{ }                  -1 \mbox{ } -1$}  }  
                                                 {  \mbox{$\mbox{ }         -1  \mbox{ } \mbox{ } \mbox{ } \mbox{ } 1 \mbox{ } -1$}  }  }
										         {  \mbox{$\mbox{ }         -1  \mbox{ }                           -1 \mbox{ } \mbox{ } \mbox{ } \mbox{ } 1$}  }  \right) \mbox{ } .  
\ee
\ni{\bf Remark 13} Then setting 
\be
0 = \mbox{det} \left( \underline{\underline{L}} - \lambda \, \underline{\underline{I}} \right) = \xi^3 - 3 \, \xi - 2 \mbox{ } , 
\ee
for
\be 
\xi := 2 - \lambda \mbox{ } ,  
\ee 
the Factor Theorem gives that $\xi = -1$ solves, reducing the problem to a quadratic equation. 
Consequently, 
\be
(\xi - 2)(\xi + 1)^2 = 0 \mbox{ } ,  
\ee 
so the eigenvalues are $\xi = 2$, i.e.\ $\lambda = 0$ with multiplicity 1, and $\xi = -1$ i.e.\ $\lambda = 3$ with multiplicity 2.

\mbox{ }

\ni{\bf Remark 14} Corresponding orthonormal eigenvectors are, respectively, 
\be 
\frac{1}{\sqrt{3}}\left( \stackrel{  \stackrel{ \mbox{1}  } 
                                              { \mbox{1}  }  }
										      {  1 } \right)                    \mbox{ } , \mbox{ } \mbox{ }
\frac{1}{\sqrt{2}}\left( \stackrel{  \stackrel{  \mbox{ }  \mbox{ } \mbox{0} }  
                                              {                     \mbox{$-1$} }  }
										      {  \mbox{ } \mbox{ }  1 } \right) \mbox{ } , \mbox{ } \mbox{ }
\frac{1}{\sqrt{6}}\left( \stackrel{  \stackrel{  \mbox{ }  \mbox{ } \mbox{2} }  
                                              {                    \mbox{$-1$} }  }
										      {                    \mbox{$-1$} } \right) \mbox{ } . 
\label{L-evectors}
\ee
\ni{\bf Remark 15} The first of these corresponds to eigenvalue 0 and is the centre of mass coordinate. 
This occurs no matter what $N$ is, and contributes nothing to the diagonalized relative Jacobi form of the inertia quadric.
Thereby, this can be considered to involve 1 coordinate vector less: $n = N - 1$ coordinate vectors, so we write it as
\be
I(\underline{\widetilde{R}}_a) = \widetilde{Y}_{aa}\widetilde{R}_a\mbox{}^2 \mbox{ } . 
\ee 
The $\widetilde{R}_i$ here are proportional to the conventional relative Jacobi coordinates. 
We tilde everything for now so as to reserve the untilded version for the conventionally used proportions themselves. 

\mbox{ } 

\ni{\bf Remark 16} For $N = 3$, 
\be
\underline{\underline{\widetilde{Y}}} = \mbox{diag}(1, \, 1) \mbox{ } , 
\ee  
so we arrive at
\be
I(\underline{\widetilde{R}}_i) = \widetilde{R}_1\mbox{}^2 + \widetilde{R}_2\mbox{}^2 \mbox{ } .
\ee 
The conventional scaling is moreover 
\be 
\uR_1 := \uq_3 - \uq_2 \mbox{ } , \mbox{ } \mbox{ } \uR_2 \mbox{ } := \mbox{ } \uq_1 - \frac{\uq_2 + \uq_3}{2} \mbox{ } ,
\label{Jac-Defs}
\ee 
which, as promised, is recognizable as consisting entirely of cluster separation vectors. 
The first is a fortiori an interparticle separation vector, whereas the second involves a 2-particle centre of mass (see Fig \ref{Jac-Med-Ineq-Fig-2}.c).  
If these are used, the diagonal relative Jacobi separation matrix $\underline{\underline{Y}}$ moreover consists of the reduced masses of the clusters in question, 
\be 
\underline{\underline{Y}} \mbox{ } = \mbox{ } \mbox{diag}\left(\frac{1}{2}, \, \frac{2}{3}\right) \mbox{ } . 
\ee 
This is indeed the standard definition of reduced mass, i.e.\ conceptually 
\be
\frac{1}{\mu} \mbox{ } = \mbox{ } \frac{1}{\mm_1} + \frac{1}{\mm_2} \mbox{ } , 
\ee 
which rearranges to the more computationally immediate form
\be
\mu \mbox{ } = \mbox{ } \frac{\mm_1\mm_2}{\mm_1 + \mm_2} \mbox{ } . 
\ee 
For equal masses, this gives 
\be
\mu_1 \mbox{ } = \mbox{ } \frac{1 \times 1}{1 + 1} \mbox{ } = \mbox{ } \frac{1}{2}
\ee 
and 
\be
\mu_2 \mbox{ } = \mbox{ } \frac{1 \times 2}{1 + 2} \mbox{ } = \mbox{ } \frac{2}{3}
\ee 
as claimed. 
Thus the relative Jacobi separation matrix can be allotted a further, now conceptual, name -- {\it reduced mass matrix} -- 
with reference to the cluster subsystems picked out in the allocation of the particular Jacobi coordinates in hand.
We mark this be replacing the $\underline{\underline{Y}}$ notation with $\underline{\underline{M}}$, 
which we take to be a capital $\mu$ standing for both `mass' and `diagonal' (in the manner that $\Lambda$ is probably the most common notation for a diagonal matrix).  
So we end up with a relative Jacobi inertia quadric of the form 
\be 
I(\underline{R}_i) = M_{ij} R_i R_j \mbox{ } . 
\ee 
\ni{\bf Remark 17} For $N = 3$, the sole ambiguity in picking out cluster subsystems in forming Jacobi coordinates is which two particles to start with. 
So there are 3 possible clustering choices, corresponding to the second orthonormal eigenvector above being free to have its zero in whichever component.\footnote{For $N \geq 4$, 
there are further ambiguities, which can be shown to result from $N \geq 4$ points supporting multiple shapes of tree graph (see e.g. \cite{I, IV}). 
Jacobi coordinates are widely used for instance in Celestial Mechanics \cite{Marchal} and in Molecular Physics \cite{LR97}.}
%
We denote the above choice by by $\uR^{(1)}$ alias $\uR^{(a)}$, 
and the clusters with $\ur_{31}$ and $\ur_{12}$ as their first relative Jacobi coordinate 
               by $\uR^{(2)}$ alias $\uR^{(b)}$ 
			  and $\uR^{(3)}$ alias $\uR^{(c)}$ respectively. 

\mbox{ }

\ni{\bf Definition 8} We furthermore denote $\mu_1$ by $\mu_{\sss}$ -- {\it side Jacobi mass}    -- 
                                        and $\mu_2$ by $\mu_{\sm}$:   {\it median Jacobi mass}.  
This is {\sl possible} since the $\mu_i$ are cluster choice independent, 
and {\sl useful} by its replacing the 1 and 2 labels with more conceptually meaningful and memorable labels, $s$ for side and $m$ for median.  
We follow suit by calling the triangle model's first and second relative Jacobi vectors the {\it side} and {\it median} vectors (for all that these {\sl are} cluster-dependent).  
I.e.\ 
\be 
R_{1}^{(a_i)} = a_i \mbox{ } , 
\ee 
\be
R_{2}^{(a_i)} = m_i \mbox{ } . 
\ee 
\ni{\bf Corollary 5} i) 
\be
R_2^{(i)\,2} \mbox{ } = \mbox{ } \frac{3}{4} \, J_{ij} R_1^{(i)\,2} \mbox{ } , 
\label{1}
\ee
inverting to ii) 
\be
R_1^{(i)\,2} \mbox{ } = \mbox{ } \frac{4}{3} \, J_{ij} R_2^{(i)\,2} \mbox{ } . 
\label{2}
\ee
{\underline{Proof}} Substitute (\ref{Jac-Defs}) into the Linear Algebra form of the sides--medians relation (\ref{0.8}). $\Box$

\section{Mass-weighted Jacobi coordinates}

\ni Our principal interest in the current paper is moreover in notions deriving from the following. 

\mbox{ }

\ni{\bf Structure 2} {\it Mass-weighted relative Jacobi coordinates} are given by   
\beq
\urho_a := \sqrt{\mu_a} \uR_{a} \mbox{ } , 
\label{rho-R}
\eeq 
where the $a$-index takes values 1 and 2.  

\mbox{ }

\ni{\bf Structure 3} {\it Mass-weighted relative Jacobi separations} are the magnitudes of the preceding, 
\beq
\rho_a := \sqrt{\mu_a} R_a 
\eeq 
\ni{\bf Remark 18} Thus computationally, 
\beq
\rho_1^{(a)} := \sqrt{\mu_1 }R_1^{(a)} \mbox{ } = \mbox{ } \frac{a}{\sqrt{2}}  \mbox{ } \mbox{ and cycles } , 
\eeq 
alongside 
\beq
\rho_1^{(a)} := \sqrt{\mu_1}R_1^{(a)} \mbox{ } = \mbox{ } \frac{m_a}{\sqrt{2}}  \mbox{ } = \mbox{ } \frac{\sqrt{2 \, b^2 + 2 \, c^2 - a^2}}{2\sqrt{2}} \mbox{ } \mbox{ and cycles } .   
\eeq
\ni{\bf Definition 9} The {\it mass-weighted inertia quadric} is 
\be 
I(\urho\mbox{}_a) \mbox{ } = \mbox{ } \sum_a \rho_a\mbox{}^2 
                  \mbox{ } = \mbox{ } \rho_1\mbox{}^2 + \rho_2\mbox{}^2 \mbox{ } , 
\ee 
where the last equality is for $N = 3$. 
Computationally, this amounts to returning to the previous section's tilded formulation. 
So one motivation for the mass-weighted relative Jacobi coordinates is that they are what drops out of the Linea Algebra approach. 
Another follows from the matrix in the quadric being the identity, alongside the following interpretation. 

\mbox{ }

\ni{\bf Structure 4} {\it Relative space} is the space of independent relative separations. 
This is moreover equipped the standard flat metric. 
This is numerically equal to $\underline{\underline{\widetilde{Y}}}$, 
but merits a new conceptual name $\underline{\underline{\widetilde{R}}}$ for standing for `relative-space', and yet is, in any case, computationally just the identity matrix, 
$\underline{\underline{1}}$.

\mbox{ }

\ni{\bf Remark 19} The Cartesian equivalence in this (mass-weighted notion of) relative space of these moreover constitutes the 

\mbox{ }

\ni{\bf `Jacobian' first motivation} for placing medians on equal footing to sides.  
Namely, that mass-weighted medians and mass-weighted sides are on an identical geometrical footing in (mass-weighted) relative space. 
These are moreover what drops out of the linear algebra most directly, in obtaining Jacobi coordinates by diagonalization. 
Motivating relative Jacobi coordinates themselves has moreover further parts to it. 
For, aside from their usefulness in treating the $N$-body problem, they turn out to be coordinates in terms of which the shape space's natural coordinates are simple \cite{III}. 

\mbox{ }

\ni{\bf Remark 20} Let us next point out the further interpretation that the mass-weighted Jacobi separations are 
related to the more widely known {\it partial moments of inertia} $I_a$ by 
\be 
\rho_a = \sqrt{I_a} \mbox{ } \mbox{ i.e.\ } \mbox{ } I_a = \rho_a\mbox{}^2 \mbox{ } . 
\label{I-rho}
\ee 
In particular, with clustering labels explicit,  
\be 
I_1^{(a)} = \rho_1\mbox{}^2 = \mu_1 R_1\mbox{}^2 \mbox{ } = \mbox{ } \frac{a^2}{2} \mbox{ } \mbox{ and cycles } , 
\label{I-1}
\ee 
and 
\be 
I_2^{(a)} = \rho_2 \mbox{}^2 = \mu_2 R_2\mbox{}^2 \mbox{ } = \mbox{ } \frac{m_a^2}{2} \mbox{ } \mbox{ and cycles } .  
\label{I-2}
\ee
\ni{\bf Definition 10} More familiarly, summing over disjoint partial moments rather than over clusters, the {\it total moment of inertia} is  
\be 
I^{(a)} := I_1^{(a)} + I_2^{(a)} \mbox{ } . 
\label{Proto-Partial}
\ee 
The definition here is that the total object is the sum of all disjoint partial contributions.  

\mbox{ }

\ni{\bf Lemma 2} (Democratic formula for the moment of inertia)
\be
I \mbox{ } = \mbox{ } \frac{  a^2 + b^2 + c^2  }{  3  } \mbox{ } = \mbox{ } \frac{1}{3}\sum_i a_i^2 \mbox{ } = \mbox{ } \frac{P_{\sI\sI}}{3} \mbox{ } .   
\label{Iabc}
\ee
\ni {\underline{Proof}} 
\be
I^{(a)}           =           I_1^{(a)} + I_2^{(a)} 
         \mbox{ } = \mbox{ }  \frac{1}{2} \, a^2 + \frac{2}{3} m_a^2 
         \mbox{ } = \mbox{ }  \frac{a^2}{2} \mbox{ } + \mbox{ } \frac{2}{3} \frac{2 \, b^2 + 2 \, c^2 - a^2}{4}
         \mbox{ } = \mbox{ }  \frac{3 \, a^2 + 2 \, b^2 + 2 \, c^2 - a^2}{6} \mbox{ } ,
\ee  
from which the result follows.
The first equality is (\ref{Proto-Partial}), 
the second uses (\ref{I-1}, \ref{I-2}),  
the third uses Corollary 2, 
and the last two steps are just elementary algebraic tidying.  $\Box$

\mbox{ }

\ni{\bf Remark 21} By this Lemma's right-hand-side's democracy invariance, we are entitled to rewrite (\ref{Proto-Partial}) stripped of its left-hand side clustering dependence $(a)$: 
\be 
I := I_1^{(a)} + I_2^{(a)} \mbox{  or cycles } . 
\ee 
It is also clear from the $I$-$\rho$ inter-relation and $I(\urho_a)$ formula that total moment of inertia is another name for the inertia quadric. 
One could argue that $I(\uq_I)$ and $I(\ur_{IJ})$ were {\it a priori} clustering-independent formulations.
Whereupon, $I(\urho\mbox{}_a)$ introduced prima facie clustering dependent features. 
Further inspection, however, confirms the cluster-dependent labels on these to be spurious since labelling-independence can indeed be maintained in Jacobi coordinates.
So $I$ is inherently cluster-independent. 
Indeed rewriting this cluster-dependently in a manner that could not be further reformulated out would indicate inconsistency of procedure, which we have therby now circumvented.  

\mbox{ }

\ni{\bf Corollary 6} i) 
\be
\rho_2^{(i)\,2} = J_{ij} \rho_1^{(i)\,2} \mbox{ } , 
\label{5}
\ee
inverting to ii) 
\be
\rho_1^{(i)\,2} = J_{ij} \rho_2^{(i)\,2} \mbox{ } . 
\label{6}
\ee
iii) A slightly tidier version in terms of partial moments of inertia is
\be
I_2^{(i)} = J_{ij} I_1^{(i)} \mbox{ } , 
\label{5b}
\ee
inverting to iv) 
\be
I_1^{(i)} = J_{ij} I_2^{(i)} \mbox{ } . 
\label{6b}
\ee
{\underline{Proof}} Substitute (\ref{rho-R}) into the Linear Algebra form of the sides--medians relation (\ref{0.8}): 
\be
\frac{\rho_2^{(i) \, 2}}{\mu_2} \mbox{ } = \mbox{ } \frac{3}{4} J_{ij} \frac{\rho_1^{(i) \, 2}}{\mu_1} \mbox{ } \Rightarrow \mbox{ } \mbox{ }
                                \frac{3}{4} \times \frac{4}{3} J_{ij} \rho_i\mbox{}^2  
								\mbox{ } = \mbox{ }                       1                                                            \mbox{ }.
\ee  
Then use (\ref{I-rho}) to obtain iii) and iv). $\Box$

\mbox{ }

\ni{\bf Remark 22} Note that the mass-weighting cleans out the awkward numerical factor of $\frac{4}{3}$ in eqs (\ref{1}-\ref{2}), revealing this to be 
\be
\frac{4}{3} \mbox{ } = \mbox{ } \frac{\mu_{\sm}}{\mu_{\sss}} \mbox{ } . 
\label{4/3} 
\ee

\section{Consequent Heron--Jacobi formulae}

\ni{\bf Definition 11} Let us introduce the {\it mass-weighted semi-perimeter} denoted by $\sigma$, and the {\it mass-weighted semi-medimeter} denoted by $\varsigma$.  

\mbox{ }

\ni{\bf Remark 23} Passing to mass weighted Jacobi versions of Heron's formula requires furthermore knowing how the bounding quantities scale.  

\mbox{ }

\ni{\bf Lemma 3} i) {\it Mass-weighted semi-perimeter} is a mass-weighting side-vector,\footnote{See \cite{Jac-Med-Ineq} for further exposition of what 
`side-vector', `median-vector' and `side-median bivector' mean.}  
\be
\sigma           :=          \sqrt{\mu_{\sss}}s 
        \mbox{ }  = \mbox{ } \frac{s}{\sqrt{2}} \mbox{ } .
\ee 
ii) {\it Mass-weighted semi-medimeter} is a mass-weighting median-vector, 
\be
\varsigma          :=           \sqrt{\mu_{\sm}}s 
          \mbox{ }  = \mbox{ } \sqrt{\frac{2}{3}} \, s \mbox{ } .
\ee 
\ni iii) Area is a mass-weighting side--median bivector,
\be
\bigupalpha\mbox{rea}          =          \sqrt{\mu_{\sss}\mu_{\sm}} \, \mbox{Area} 
                      \mbox{ } = \mbox{ } \frac{   \mbox{Area}   }{   \sqrt{3}   } \mbox{ } .
\ee
\ni{\underline{Proof}} i) 
\be
s \mbox{ } := \mbox{ } \sum_i a_i 
   \mbox{ } = \mbox{ } \sum_i R_1^{(i)} 
   \mbox{ } = \mbox{ } \frac{1}{\sqrt{\mu_1}}   \sum_i \rho_1^{(i)} 
   \mbox{ } = \mbox{ } \frac{\sigma}{\sqrt{\mu_{\ms}}}
   \mbox{ } = \mbox{ } \sqrt{2} \sigma                                \mbox{ } . 
\ee 
ii)
\be
\ms \mbox{ } := \mbox{ } \sum_i m_i \sum_i R_2^{(i)} 
     \mbox{ } = \mbox{ } \frac{1}{\sqrt{\mu_2}}   \sum_i \rho_2^{(i)} 
	 \mbox{ } = \mbox{ } \frac{\varsigma}{\mu_{\sm}} 
	 \mbox{ } = \mbox{ } \sqrt{\frac{3}{2}}\varsigma                  \mbox{ } . 
\ee
iii) 
\be
\mbox{Area} \mbox{ } = \mbox{ } \frac{1}{2}\{\uR_1 \times \uR_2 \}_3 
            \mbox{ } = \mbox{ } \frac{1}{2\sqrt{\mu_1\mu_2}} \, \{\urho_1 \times \urho_2 \}_3 
			\mbox{ } = \mbox{ } \frac{\bigupalpha\mbox{rea}}{\sqrt{\mu_1\mu_2}} 
			\mbox{ } = \mbox{ } \frac{\bigupalpha\mbox{rea}}{\sqrt{\frac{1}{2} \, \frac{2}{3}}}
			\mbox{ } = \mbox{ } \sqrt{3} \, {\bigupalpha\mbox{rea}}                              \mbox{ } .  \mbox{ } \Box 
\ee 
\ni{\bf Theorem 4} (Mass-weighted area's {\it Heron--Jacobi formulae} in terms of each of mass-weighted sides and mass-weighted medians) 
\be 
\sqrt{    \sigma     \left(  \sigma - \rho_1^{(a)}  \right)   
                     \left(  \sigma - \rho_1^{(b)}  \right)   
					 \left(  \sigma - \rho_1^{(c)}  \right)   } 
\mbox{ } = \bigupalpha\mbox{rea}  = \mbox{ } 
\sqrt{    \varsigma  \left(  \varsigma - \rho_2^{(a)}  \right)   
                     \left(  \varsigma - \rho_2^{(b)}  \right)   
					 \left(  \varsigma - \rho_2^{(c)}  \right)   }  \mbox{ } .
					 \label{HHF}
\ee
\ni{\underline{Proof}} See Fig \ref{mw-Heron}. $\Box$
%
{            \begin{figure}[!ht]
\centering
\includegraphics[width=0.8\textwidth]{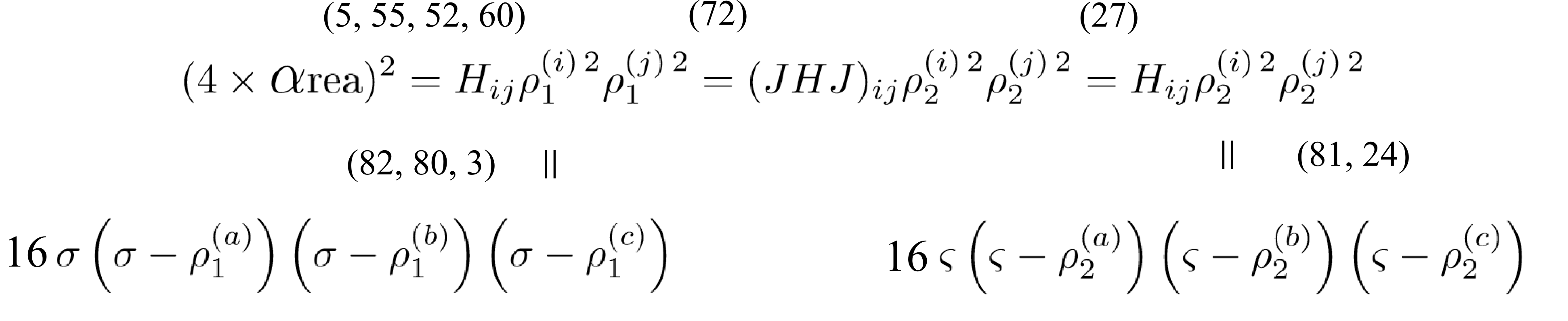}
\caption[Text der im Bilderverzeichnis auftaucht]{        \footnotesize{Proof of Theorem 4.  }  } 
\label{mw-Heron} \end{figure}           }

\ni{\bf Remark 24} Note that the sides and medians versions are now on an equal footing: 
without any constant prefactor difference like the $\frac{4}{3}$ in the mass-unweighted medians case (Theorem 1) relative to the mass-unweighted sides case (Theorem 2).
This amounts to explaining the $\frac{4}{3}$ factor discrepancy between the medians' Heron's formula and the standard sides' Heron's formula 
as resulting from formulating Heron's formula for area rather than for mass-weighted area.
Enter the Jacobi picture, then mass-weighted area is more natural for the formulation of Heron's formulae, 
since now the mass-weighted sides' version and the mass-weighted medians' versions are on an identical footing without such a numerical factor. 
The numerical factor's significance is thus unmasked to be the same ratio of reduced masses as tidied up the preceding Linear Algebra, as per eq (\ref{4/3}). 
This is embodied in the Heron--Jacobi formulae (\ref{HHF}), to be compared with the less symmetrical, and thus less transparent mass-unweighted summary equation (\ref{Summary-1}).

\mbox{ }

\ni{\bf Corollary 7} (Mass weighted area's mass-weighted {\it sides-and-medians-democratic Heron--Jacobi formula})
\be 
\bigupalpha\mbox{rea} \mbox{ } = \mbox{ } \sqrt{   \frac{1}{2} \sum_{i = 1}^2  
                                                   \sigma_i  \left(  \sigma_i - \rho_i^{(a)}  \right) 
						          \left(  \sigma_i - \rho_i^{(b)}  \right)
								  \left(  \sigma_i - \rho_i^{(c)}  \right)    } 
\ee 
where $\sigma_1 := \sigma$ and $\sigma_2 := \varsigma$. 

\mbox{ }

\ni{\underline{Proof}} Add the squares of the two equalities and then square-root. $\Box$

\mbox{ }

\ni{\bf Remark 25} This is the {\sl root mean sum} of the sides' and medians' Heron--Jacobi formulae.

\section{Diagonalizing the Heron matrix gives the Hopf Map and Kendall's Theorem}\label{HoKe}

{\bf Remark 26} Set 
\be
0 = \mbox{det}\left(\underline{\underline{H}} - \lambda \underline{\underline{I}} \right) = \nu^3 - \nu + 2 \mbox{ } , 
\ee
for
\be 
\nu := - 1 - \lambda \mbox{ } . 
\ee 
Then the Factor Theorem gives that $\nu = - 2$ solves, reducing the problem to a quadratic equation. 
Consequently, 
\be
(\nu - 1)^2(\nu + 2) = 0 
\ee 
so the eigenvalues are $\nu = - 2$, i.e.\ $\lambda = 1$ with multiplicity 1, and $\nu = 1$, i.e.\ $\lambda = -2$ with multiplicity 2.

\mbox{ }

\ni{\bf Remark 27} Corresponding orthonormal eigenvectors are, respectively, 
\be 
\frac{1}{\sqrt{3}}    \left(    \stackrel{           \mbox{1}  }{  \stackrel{  \mbox{1}     }{  1  }  }  \right)  \mbox{ } , \mbox{ } \mbox{ } 
\frac{1}{\sqrt{2}}    \left(    \stackrel{  \mbox{ } \mbox{1}  }{  \stackrel{  \mbox{$-1$}  }{ \mbox{ } 0  }  }  \right)  \mbox{ } , \mbox{ } \mbox{ } 
\frac{1}{\sqrt{6}}    \left(    \stackrel{  \mbox{ } \mbox{1}  }{  \stackrel{  \mbox{ } \mbox{1}     }{ -2  }  }  \right)  \mbox{ } .  
\label{H-evectors}
\ee 
\ni{\bf Structure 5} The diagonalizing variables are thus 
\be
\overline{a}^2 \mbox{ } = \mbox{ } \frac{a^2 + b^2 + c^2}{\sqrt{3}} \mbox{ } , 
\ee 
\be
\overline{b}^2 \mbox{ } = \mbox{ } \frac{a^2 - b^2}{\sqrt{2}} \mbox{ } , \mbox{ } \mbox{ and }
\ee 
\be
\overline{c}^2 \mbox{ } = \mbox{ } \frac{a^2 + b^2 - 2 \, c^2}{\sqrt{2}} \mbox{ } .   
\ee 
\ni{\bf Remark 28} The Heron quadratic form (\ref{Heron-form}) is hence equal to 
\be
\sum_i \Lambda_{ii} \overline{a}_i\mbox{}^2 \overline{a}_i\mbox{}^2 
\ee 
for {\it diagonalized Heron matrix}
\be 
\Lambda_{ij} = \mbox{diag}(1, \, - 2, \, - 2) \mbox{ } , 
\ee 
which related to the original Heron matrix $H_{ij}$ by conjugation with the $P_{ij}$ formed by using (\ref{H-evectors}) orthonormal eigenvectors as its columns.  
Thus we have derived that Heron's formula also takes the following form. 

\mbox{ }

\ni{\bf Theorem 5 (Diagonal Heron formula)} 
\be 
Area = \frac{1}{4} \, \sqrt{  \overline{a}\mbox{}^4 - 2(\overline{b}\mbox{}^2 + \overline{c}\mbox{}^2)  } \mbox{ } .  
\ee 
\ni {\bf Corollary 8} Next multiply both sides through by $\frac{4}{\overline{a}_i\mbox{}^2}$ to obtain 
\be 
\frac{4 \times Area}{\overline{a}^2} = \sqrt{  1 - \left[    \sqrt{2}\left(    \frac{  \overline{b} }{  \overline{a}  }    \right)^2    \right]^2 
                                                 - \left[    \sqrt{2}\left(    \frac{  \overline{c} }{  \overline{a}  }    \right)^2    \right]^2    } \mbox{ } .  
\ee
\ni{\bf Remark 29} The $4 \times Area$ scaling present in the Hopf quantity has its 4 come from 
\be
Area = \frac{1}{4}\sqrt{(\mbox{expanded Heron form})} \mbox{ } ,
\ee 
so 
\be
(4 \times Area)^2 = (\mbox{expanded Heron form}) \mbox{ } . 
\ee 
\ni{\bf Definition 12} It is thus natural to finally define the rescaled {\it ratio variables} 
\be 
Z \mbox{ } := \mbox{ } \sqrt{2}  \left(  \frac{  \overline{c}  }{  \overline{a}  }  \right)^2 
  \mbox{ }  = \mbox{ } \frac{  a^2 + b^2 - 2 \, c^2  }{  a^2 + b^2 + c^2 }                         \mbox{ } .  
\label{Z}
\ee 
\be 
X \mbox{ } := \mbox{ } \sqrt{2}  \left(  \frac{\overline{b}}{\overline{a}}  \right)^2 
  \mbox{ }  = \mbox{ } \frac{\sqrt{3}(a^2 - b^2)  }{  a^2 + b^2 + c^2 }                            \mbox{ } \mbox{ and } 
\label{X}
\ee 
\be 
Y \mbox{ }:= \mbox{ } \frac{4 \times Area}{\overline{a}^2}                                         \mbox{ } . 
\ee 
\ni{\bf Remark 30} The denominator of the ratio is proportional to the moment of inertia by Lemma 2. 
$Y$ is moreover to be interpreted precisely as {\it mass-weighted area per unit moment of inertia}.

\mbox{ }

\ni{\bf Remark 31} In terms of these rescaled ratio variables, Heron's formula has been reduced to just the following.

\mbox{ }

\ni{\bf Corollary 9} The rescaled ratio variables version of the diagonal Heron formula is  
\be 
X^2 + Y^2 + Z^2 = 1 \mbox{ } ,  
\ee 
which is mathematically just the on 2-sphere condition.  

\mbox{ }

\ni{\bf Remark 32} We furthermore identify $Y = 4 \times (\mbox{mass-weighted area})$ as a Hopf quantity.

\mbox{ }

\ni{\bf Remark 33} $X$ and $Z$ are also Hopf quantities, which, in the triangle context, can moreover be interpreted as follows \cite{+Tri, FileR, III}. 
Without normalizing, one has 
\be 
Aniso \mbox{ } = \mbox{ } \frac{a^2 - b^2}{\sqrt{3}} \mbox{ } , 
\ee 
\be 
Ellip \mbox{ } = \mbox{ } \frac{a^2 + b^2 - 2 \, c^2  }{  3  } \mbox{ } . 
\ee 
One can readily check that these and $4 \times \bigupalpha rea$ obey 
\be
Aniso^2 + Ellip^2 + (4 \times \bigupalpha rea)^2 = I^2 \mbox{ } . 
\ee 
\ni{\bf Remark 34} Or, at the level of shape quantities, i.e.\ with normalization, one has  
\be 
ellip \mbox{ } = \mbox{ } \frac{ a^2 + b^2 - 2 \, c^2  }{  a^2 + b^2 + c^2  } \mbox{ } , 
\ee 
\be
aniso \mbox{ } = \mbox{ } \frac{  a^2 - b^2  }{  \sqrt{3} \{ a^2 + b^2 + c^2 \}  } \mbox{ } .  
\ee
One can also check that these and $4 \times area$ obey the on-sphere condition 
\be
aniso^2 + ellip^2 + (4 \times \upalpha rea)^2 = 1 \mbox{ } .  
\ee 
for normalized mass-weighted area 
\be 
\upalpha rea :=  \frac{\bigupalpha rea}{I} \mbox{ } .  
\ee
\ni{\bf Remark 35} Anisoscelesness and ellipticity can moroever now be interpreted as two of the eigenvectors of the Heron map $\underline{\underline{H}}$.
 
\mbox{ }

\ni{\bf Remark 36} Moreover, due to $\underline{\underline{H}}$ and $\underline{\underline{J}}$ commuting with each other, these maps share their eigenvectors.
[This is  well-known in Quantum Mechanics, under the name of `complete set of commuting observables' (CSCO), and in Methods of Mathematical Physics.] 
Thus anisoscelesness and ellipticity are also eigenvectors of the sides--medians involution $\underline{\underline{J}}$.

\section{Median--sides interchange form invariance of diagonal Heron--Hopf formula}

\ni{\bf Corollary 10} In terms of the medians, i) 
\be 
Ellip = -\frac{4}{3}\frac{m_a\mbox{}^2 + m_b\mbox{}^2 - 2 \, m_c\mbox{}^2  }{3} \mbox{ } , 
\ee 
\be
Aniso = - \frac{4}{3}\frac{m_a\mbox{}^2 - m_b\mbox{}^2}{\sqrt{3}} \mbox{ } . 
\ee 
ii) at the level of shape quantities, 
\be 
ellip = - \frac{m_a\mbox{}^2 + m_b\mbox{}^2 - 2 \, m_c\mbox{}^2  }{  m_a\mbox{}^2 + m_b\mbox{}^2 + m_c\mbox{}^2  } \mbox{ } , 
\ee 
\be
aniso = - \frac{  m_a\mbox{}^2 - m_b\mbox{}^2  }{\sqrt{3} \{ m_a\mbox{}^2 + m_b\mbox{}^2 + m_c\mbox{}^2 \} } \mbox{ } . 
\ee
\ni{\underline{Proof}} Use the below Lemma and the sides--medians involution. $\Box$ 

\mbox{ }

\ni{\bf Lemma 4} (Democratic medians form of total moment of inertia)
\be 
I \mbox{ } = \mbox{ } \frac{4}{9} \sum_i m_i\mbox{}^2 \mbox{ } = \mbox{ }  \frac{4}{9}M_{\sI\sI} \mbox{ } .  
\ee
\ni{\underline{Proof}} As in the proof of Lemma 2,
\be 
I \mbox{ } = \mbox{ } \frac{1}{2} \, a^2 + \frac{2}{3} \, m_a\mbox{}^2 \mbox{ } , 
\ee 
but now substitute for $a^2$ using the sides-to-medians involution, 
\be
I \mbox{ } = \mbox{ } \frac{1}{2} \, \frac{4}{9} (2 \, m_b\mbox{}^2 + 2 \, m_c\mbox{}^2 - m_a\mbox{}^2 ) + \frac{2}{3}m_a\mbox{}^2 
  \mbox{ } = \mbox{ } \frac{4}{9}(m_a\mbox{}^2 + m_b\mbox{}^2 + m_c\mbox{}^2) 
  \mbox{ } = \mbox{ } \frac{4}{9} \sum_i m_i\mbox{}^2                                                                            
  \mbox{ } = \mbox{ } \frac{4}{9} \, M_{\sI\sI}                                                                                          \mbox{ } ,  
\ee 
as desired, using Definition (\ref{MII}) in the last step. $\Box$

\mbox{  }

\ni{\bf Remark 37} Expressing $ellip$ and $aniso$ in terms of medians instead does not moreover affect the diagonality.  
It does flip the signs over, but this is part and parcel of the allowed conventions in setting up a Cartesian axis system.  
The Heron--Hopf formula is thus independent of whether one is conceiving in terms of sides or of medians; 
the Hopf quantities offer a third point of view that is side--median symmetric.  This constitutes the 

\mbox{ }

\ni{\bf `Hopfian' second motivation} for treating medians on the same footing as sides.

\section{Further Linear Algebra, now of Hopf Quantities}

\ni{\bf Structure 6} Let us further formulate anisoscelesness and ellipticity in Linear Algebra terms as follows.  
\be
Aniso = {\cal A}_i a_i\mbox{}^2 = {\cal A}_i J_{ij} m_i\mbox{}^2 = - {\cal A}_i m_i\mbox{}^2 \mbox{ }   
\ee 
and 
\be
Ellip = {\cal E}_i a_i\mbox{}^2 = {\cal E}_i J_{ij} m_i\mbox{}^2 = - {\cal E}_i m_i\mbox{}^2 \mbox{ } .  
\ee 
for vectors 
\be
\underline{\cal A} := \left( \stackrel{\mbox{1}}{\stackrel{\mbox{$-1$}}{0}} \right) \mbox{ and } \mbox{ } 
\underline{\cal E} := \left( \stackrel{\mbox{1}}{\stackrel{\mbox{$1$}}{-2}} \right) \mbox{ } .
\ee 
\ni{\bf Structure 7} Introduce furthermore $\underline{\underline{E}}$ and $\underline{\underline{A}}$ matrices for ellipticity squared and anisoscelesness squared,
\be
Ellip^2 = E_{ij} a_i\mbox{}^2 a_j\mbox{}^2 \mbox{ } 
\ee 
and 
\be 
Aniso^2 = A_{ij} a_i\mbox{}^2 a_j\mbox{}^2 \mbox{ } , 
\ee 
for 
\be
\underline{\underline{E}} \mbox{ } := \mbox{ } \left( \stackrel{  \stackrel{  \mbox{$4  \mbox{ }                   -2 \mbox{ }                  -2$}  }  
                                                                 {  \mbox{$          2  \mbox{ } \mbox{ } \mbox{ }  1 \mbox{ } \mbox{ } \mbox{ } 1$}  }  }
										                         {  \mbox{$          2  \mbox{ } \mbox{ } \mbox{ }  1 \mbox{ } \mbox{ } \mbox{ } 1$}  }    \right) \mbox{ } , \mbox{ and } \mbox{ } 
\underline{\underline{A}} \mbox{ } := \mbox{ } \left( \stackrel{  \stackrel{  \mbox{$0  \mbox{ } \mbox{ } \mbox{ }   0 \mbox{ }  \mbox{ }  \mbox{ }           0$}  }  
                                                                 {  \mbox{$          0  \mbox{ } \mbox{ } \mbox{ }   1 \mbox{ }          -1$}  }  }
										                         {  \mbox{$          0  \mbox{ }          -1 \mbox{ } \mbox{ }  \mbox{ }  1$}  }     \right) \mbox{ } . 																 
\ee
Motivation for doing this for the squares is that it is these which are on a Hopf bundle-theoretic par with the Heron map $\underline{\underline{H}}$ of Sec 2. 
We then observe the following remarkable commutativity theorem. 

\mbox{ }

\ni{\bf Theorem 6} All three of the $Hopf^2$ quantities' matrices commute 

\mbox{ }

\ni i) with each other, 
\be
\mbox{\bf [} \underline{\underline{H}} \mbox{\bf ,} \, \underline{\underline{E}} \mbox{\bf ]} \mbox{ } = \mbox{ }   
\mbox{\bf [} \underline{\underline{H}} \mbox{\bf ,} \, \underline{\underline{A}} \mbox{\bf ]} \mbox{ } = \mbox{ }
\mbox{\bf [} \underline{\underline{E}} \mbox{\bf ,} \, \underline{\underline{A}} \mbox{\bf ]} \mbox{ } = \mbox{ } 0 \mbox{ } ,  \mbox{ and }
\ee 
\ni ii) with the sides-median involution $\underline{\underline{J}}$, 
\be
\mbox{\bf [} \underline{\underline{H}} \mbox{\bf ,} \, \underline{\underline{J}} \mbox{\bf ]} \mbox{ } = \mbox{ }   
\mbox{\bf [} \underline{\underline{E}} \mbox{\bf ,} \, \underline{\underline{J}} \mbox{\bf ]} \mbox{ } = \mbox{ } 
\mbox{\bf [} \underline{\underline{A}} \mbox{\bf ,} \, \underline{\underline{J}} \mbox{\bf ]} \mbox{ } = \mbox{ } 0 \mbox{ } .    
\ee 
\ni{\underline{Proof}} Use Lemma 5 below, itself established by mere matrix multiplication and the definitions of  
$\underline{\underline{E}}$, $\underline{\underline{A}}$ and eq. (\ref{K})'s $\underline{\underline{K}}$.  $\Box$ 

\mbox{ }

\ni{\bf Lemma 5}
\be
\underline{\underline{H}} \, \underline{\underline{A}} = 2 \, \underline{\underline{A}} =  \underline{\underline{A}}  \, \underline{\underline{H}}  \mbox{ } ,
\ee 
\be
\underline{\underline{H}} \, \underline{\underline{E}} = - 2 \, \underline{\underline{E}} = \underline{\underline{E}} \, \underline{\underline{H}}  \mbox{ } , 
\ee 
\be 
\underline{\underline{A}}  \, \underline{\underline{E}} = 0 = \underline{\underline{E}}  \, \underline{\underline{A}}                               \mbox{ } ,
\ee 
\be 
\underline{\underline{H}} \, \underline{\underline{J}} = \underline{\underline{K}} = \underline{\underline{J}} \, \underline{\underline{H}}         \mbox{ } ,
\ee 
\be 
\underline{\underline{J}}  \, \underline{\underline{E}} = - \underline{\underline{E}} = \underline{\underline{E}} \, \underline{\underline{J}}      \mbox{ } , 
\ee 
\be 
\underline{\underline{J}} \, \underline{\underline{A}} = - \underline{\underline{E}} = \underline{\underline{A}}  \, \underline{\underline{J}}      \mbox{ } . 
\ee 
\ni{\bf Corollary 11} (Matrix form of the Hopf on-sphere condition)
\be 
\underline{\underline{H}} + \underline{\underline{A}} + \underline{\underline{E}} \mbox{ } = \mbox{ } \frac{1}{9}  \, \underline{\underline{I}}     \mbox{ } ,  
\ee 
where \underline{\underline{I}} is here the degenerate `all unit entries matrix' 
\be 
\underline{\underline{I}} := \mbox{ } \left( \stackrel{  \stackrel{  \mbox{$1  \mbox{ } \mbox{ } 1 \mbox{ } \mbox{ } 1$}  }  
                                          {  \mbox{$1  \mbox{ } \mbox{ } 1 \mbox{ } \mbox{ } 1$}  }  }
										  {  \mbox{$1  \mbox{ } \mbox{ } 1 \mbox{ } \mbox{ } 1$}  }     \right) \mbox{ } . 
\ee 
so that $\frac{1}{9} \, \underline{\underline{I}}$ is what occurs in the moment of inertia squared regarded as a quadratic form of squares.

\section{Heron--Hopf--Kendall, Heron--Hopf and two concomitant formulae}

Let us conclude the previous three sections as follows.
Sec \ref{HoKe}'s workings readily imply the following Theorems.     

\mbox{ }

\ni{\bf Theorem 7} The diagonalized form of the mass-weighted Heron formula is 
\be
4 \times \upalpha rea = \sqrt{1 - aniso^2 - ellip^2} \mbox{ } . 
\label{HH}
\ee
This `{\it Heron--Hopf}' formula is moreover sides--to-medians symmetric.

\mbox{ }

\ni{\bf Theorem 8} The most symmetrical presentation of the diagonalized mass-weighted Heron formula is 
\be 
(4 \times \upalpha rea)^2 + aniso^2 + ellip^2 = 1 
\ee 
This `{\it Heron--Hopf--Kendall}' formula is mathematically just the on-sphere condition; 
moreover obsrving this amounts to a {\sl recovery} of {\it Kendall's Theorem} that the shape space of all triangles in 2-$d$ is a sphere (Appendix B). 

\mbox{ }

\ni{\bf Remark 38} As considered in current paper, Kendall's Theorem is in the context of distinctly labelled point-or-particle vertices without mirror image identification. 
See \cite{I, III} as regards the outcome of varying these modelling assumptions. 
A conceptual name for this theorem is `{\it triangleland sphere theorem}', whereas one for the more general Kendall theorem is `{\it N-a-gonland complex projective space theorem}'.

\mbox{ }

\ni{\bf Remark 39} The Heron--Hopf formula, as the area-subject Hopf formula in terms of ellipticity and anisoscelesness data, now has two concomitant formulae in many senses.
These are, firstly, the ellipticity-subject Hopf formula in terms of anisoscelesness and area data, 
\be
ellip = \sqrt{1 - aniso^2 - (4 \times \upalpha rea)^2} \mbox{ } . 
\ee 
Secondly, the anisoscelesness-subject Hopf formula in terms of elliptiity and area data,  
\be
aniso = \sqrt{1 - ellip^2 - (4 \times \upalpha rea)^2} \mbox{ } .   
\ee
The sense in which Heron--Hopf has a further quality than these two other formulae is that 
it is the only one among these which has a democratic i.e.\ clustering-independent \cite{LR97, III} subject.

\section{Conclusion}

\ni In the current paper, we considered placing medians of triangles on the same footing as their sides due to Jacobi and Hopf motivation, 
both of these being structures entering the Shape Theory of the space af triangles.  

\mbox{ }

\ni We first reformulated the medians--sides inter-relation as an involution $\underline{\underline{J}}$.
We observed that $\underline{\underline{J}}$(s) factor of $\frac{4}{3}$ has the same origin as the $\frac{4}{3}$ factor 
discrepancy between sides and medians in the standard Heron's formulae in space. 
Moreover, in both the involution $\underline{\underline{J}}$ and Heron's formula, 
this factor of $\frac{4}{3}$ is removed by the passage to the mass-weighted Jacobi coordinate version of the inter-relation and of the Heron's formulae.  
Thus the mass-weighted sides and mass-weighted medians versions of Heron's formula for the now also mass-weighted area are placed on an indentical footing. 
We term these the {\it Heron--Jacobi formulae}. 
In the process, the orignal factor of $\frac{4}{3}$ is identified to be the ratio of the medians' Jacobi mass to the sides' Jacobi mass.  

\mbox{ } 

\ni Secondly, we point to the elsewise well-known Hopf coordinates diagonalizing the Heron map $\underline{\underline{H}}$, a fact that appears to have hitherto escaped attention.  
Indeed, in this manner, diagonalizing Heron's map $\underline{\underline{H}}$ provides us with a new derivation both of the Hopf map, 
and of the shape space formed by the triangles being a sphere equipped with the standard round metric: Kendall's Theorem.  
What occurs at the level of the Hopf formulation of the triangle is that Heron's formula has become a `{\it Heron--Hopf}' formula that is one and the same as 
the on-sphere condition determining that the shape space of triangles is a sphere.  

\mbox{ }

\ni The factor of 4 in the Hopf quantity that, in its 3-body problem incarnation, is the `tetra-area' is moreover accounted for in the current paper's working 
as being none other than the prefactor of 1/4 in the expanded version of Heron's formula.  

\mbox{ }

\ni The other two Hopf quantities are, in their 3-body problem incarnation, to be interpreted as ellipticity and anisoscelesness.
In the current paper, these now receive the further enlightening interpretation of being Heron map $\underline{\underline{H}}$ eigenvectors, 
                               which are moreover also sides--medians involution $\underline{\underline{J}}$ eigenvectors by 
the commutation relation between $\underline{\underline{H}}$ and $\underline{\underline{J}}$ also established in this paper.  

\mbox{ }

\ni Our {\it Heron--Hopf formula} is obtained by diagonalizing the expanded Heron form, keeping mass-weighted area as the subject.
This is moreover mathematically in the form of an on-sphere condition, which, if represented symmetrically, we term the {\it Heron--Hopf--Kendall formula} 
in honour of Kendall's iconic shape sphere of triangles.  
We finally argued that one can (almost) just as well interpret ellipticity or anisoscelesness as the subject, giving two further concomitants of the Heron--Hopf formula.  

\mbox{ }

\ni{\bf Acknowledgments} I thank Chris Isham and Don Page for previous discussions, 
                                 and 
                                 Malcolm MacCallum, Reza Tavakol, Jeremy Butterfield and Enrique Alvarez for support with my career. 

\vspace{10in}
								 
\begin{appendices}
								 
\section{The Hopf map}\label{App-1}
%
{            \begin{figure}[!ht]
\centering
\includegraphics[width=0.2\textwidth]{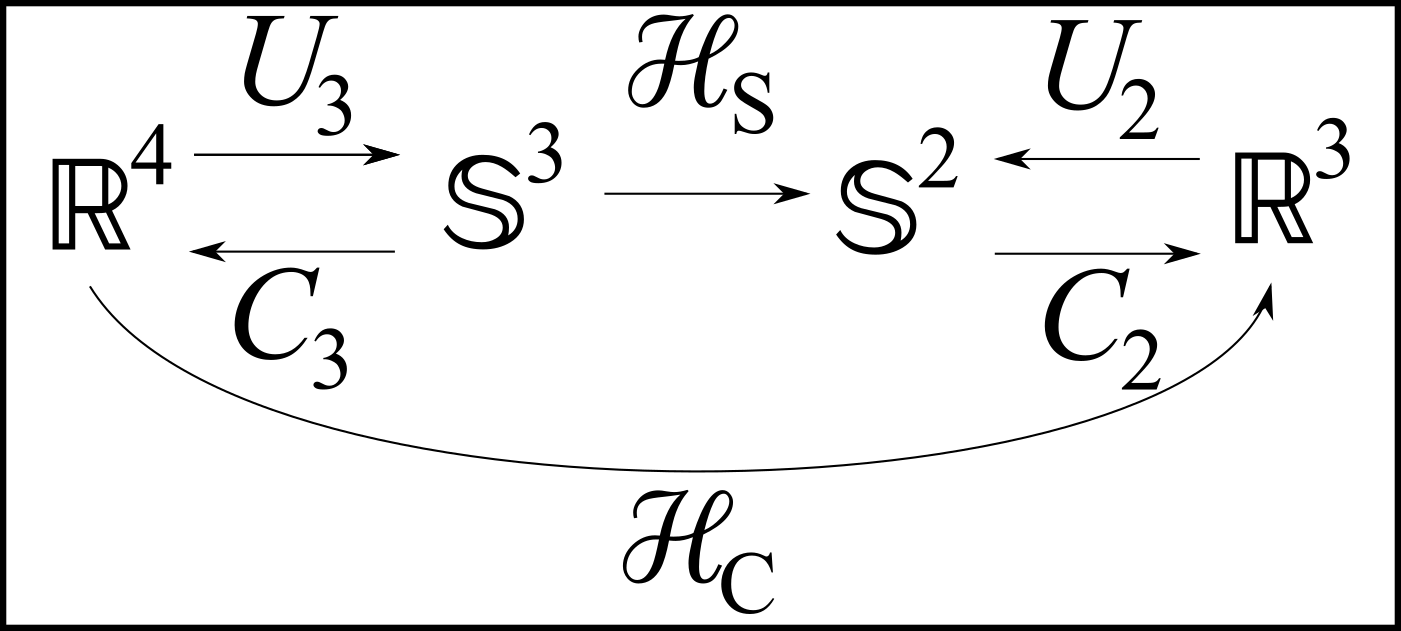}
\caption[Text der im Bilderverzeichnis auftaucht]{        \footnotesize{Hopf map ${\cal H}_{\sS}$ from $\mathbb{S}^3$ to $\mathbb{S}^2$, 
including also mapping the natural ambient $\mathbb{R}^4$ for the $\mathbb{S}^3$ to a less obviously realized natural ambient $\mathbb{R}^3$ for the $\mathbb{S}^2$ 
by the Hopf Cartesian map ${\cal H}_{\sC}$. 
$U_k$ are here unit vector maps to the $k$-sphere, and $C_k$ are cone maps from the $k$-sphere.}  } 
\label{Hopf-Map} \end{figure}           }

\ni{\bf Structure 8} The main such considered in the current paper is the simplest case, which we start to outline in Fig \ref{Hopf-Map}.

\mbox{ }

\ni{\bf Structure 9} The $\mathbb{R}^3$'s Cartesian directions moreover make equable use of the $\mathbb{R}^4$'s in the following way 
(modulo signs and permutations as regards which Hopf quantities are suffixed $X$, $Y$ and $Z$).  
\be
Hopf_X := 2 \, \ux \cdot \ux^{\prime} 
\mbox{ } ,
\ee
\be
Hopf_Y := 2 (\ux \cr \ux^{\prime})_3 
\mbox{ } ,
\ee
and 
\be
Hopf_Z := x^2 - x^{\prime \, 2} 
\mbox{ } .
\ee
\ni{\bf Structure 10} Normalizing, the unit Cartesian directions in $\mathbb{R}^3$ are 
\be 
hopf_i = \frac{Hopf_i}{x^2 + x^{\prime \, 2}} \mbox{ } . 
\ee 
These can be readily checked to obey the on-2-sphere condition,  
\be 
\sum_{i = 1}^3 hopf_i\mbox{}^2 = 1 \mbox{ } . 
\ee  
\ni{\bf Remark 41} The above Hopf Cartesian coordinates map, ${\cal H}_{\sC}$ is related to the Hopf spheres map ${\cal H}_{\sS}$ by the composition of maps  
\be
{\cal H}_{\sC} = U_3 \, \circ \,  {\cal H}_{\sS} \, \circ \, C_2 \mbox{ } ,   
\ee
as given in Fig \ref{Hopf-Map}. 

\mbox{ }

\ni{\bf Structure 11} The Hopf spheres map can moreover be regarded as a principal fibre bundle with base space $\mathbb{S}^2$, fibre $\mathbb{S}^1 = U(1)$ structure group,  
and total space $\mathbb{S}^3$.  

\mbox{ }

\ni{\bf Remark 43} Applications of this Hopf mathematics include the following. 

\mbox{ }

\ni{\bf Application 1)} It provides a simple nontrivial example of fibre bundle and of fibration \cite{Nakahara, Frankel}.

\mbox{ }

\ni{\bf Application 2)} It is theoretically realized in space by the Dirac monopole \cite{Dir-Mon}.

\mbox{ }

\ni{\bf Application 3)}  It is realized in configuration space in study of the 3-body problem (alongside variants reviewed in \cite{III, A-Monopoles}); 
this realization is moreover shape-theoretic and so is further outlined in Appendix \ref{App-2}. 

\mbox{ }

\ni{\bf Application 4)}  It extends to two other special-dimensional cases as supported by the quaternions and octonians \cite{Husemoller}. 

\mbox{ }

\ni{\bf Application 5)}  It extends systematically to inter-relate spheres of odd dimension to complex-projective spaces (outlined in Appendix \ref{App-3}).

\mbox{ }

\ni{\bf Application 6)} Moreover Application 5) is realized in configuration space in study of the planar $N$-body problem, in what is again a shape-theoretic realization 
as outlined in Appendix \ref{App-4}.  

\vspace{10in}
	
\section{Shape-theoretic realization of the Hopf map, and Kendall's Theorem}\label{App-2}
%
{            \begin{figure}[!ht]
\centering
\includegraphics[width=0.55\textwidth]{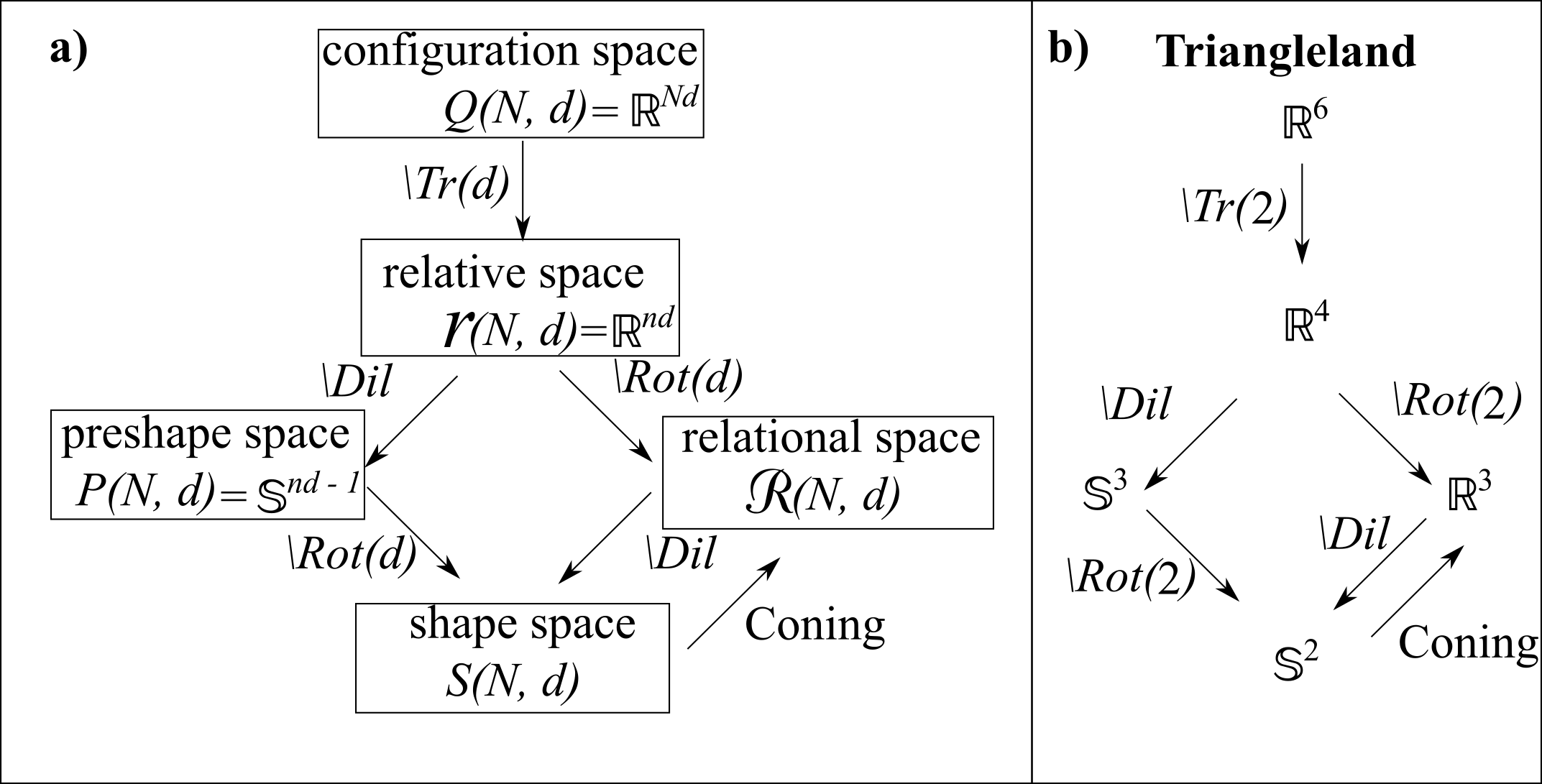}
\caption[Text der im Bilderverzeichnis auftaucht]{        \footnotesize{a) Successively more reduced notions of configuration space for Kendall's similarity Shaper Theory. 
$Tr(d)$ are translations, $Rot(d)$ are rotations and $Dil$ are dilations. 

\mbox{ }

\ni b) The more specific case of configuration spaces for the triangle in 2-$d$, in the case with with vertices distinctly labelled and mirror images held to be distinct.
Of course, now $Rot(2) = SO(2) = U(1)$, so the bottom-left quotienting is Appendix A's Hopf map ${\cal H}_S$; ${\cal H}_C$ is also realized in this diagram.}  } 
\label{Triangleland} \end{figure}           }

\ni The previous Appendix's $\ux$ and $\ux^{\prime}$ are now realized by the mass-weighted relative Jacobi vectors $\urho_1$ and $\urho_2$.  

\mbox{ }

\ni{\bf Theorem 9 (Kendall's Theorem)} The shape space of triangles, in the sense of labelled mirror-image-distinct 3-point constellations in the plane, is topologically a sphere 
and equipped furthermore at the metric level with the standard round metric.

\mbox{ }

\ni{\bf Remark 42} This can be proven geometrically \cite{Kendall84, Kendall89, Kendall}, 
by reduction of the corresponding mechanical theory \cite{FORD}, or indeed from the Hopf map \cite{III}, or, indeed, as per the current paper, 
by reformulating Heron's formula to obtain the Hopf on-sphere condition.  

\mbox{}

\ni{\bf Remark 43} See Fig \ref{S(3, 2)-Intro} for a sketch of some features of the shape sphere of triangles and \cite{III} for further details.  

\mbox{ }

\ni{\bf Remark 44} See \cite{I, III, A-Monopoles} for further unlabelled and/or mirror-image identified and/or 3-$d$ versions.
%
{            \begin{figure}[!ht]
\centering
\includegraphics[width=0.45\textwidth]{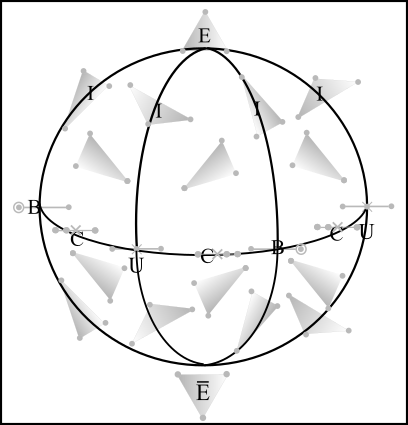}
\caption[Text der im Bilderverzeichnis auftaucht]{        \footnotesize{The triangleland sphere \cite{Kendall89, +Tri, FileR, III}.     
Equilateral triangles E are at its poles, whereas collinear configurations C form its equator.
There are 3 bimeridians of isoscelesness I corresponding to 3 labelling choices for the vertices.
$C \, \cap I$ gives 3 binary collisions B and 3 uniform collinear shapes U.} }
\label{S(3, 2)-Intro} \end{figure}          }

\vspace{10in}

\section{Generalized Hopf map for odd-dimensional spheres}\label{App-3}
%
{            \begin{figure}[!ht]
\centering
\includegraphics[width=0.4\textwidth]{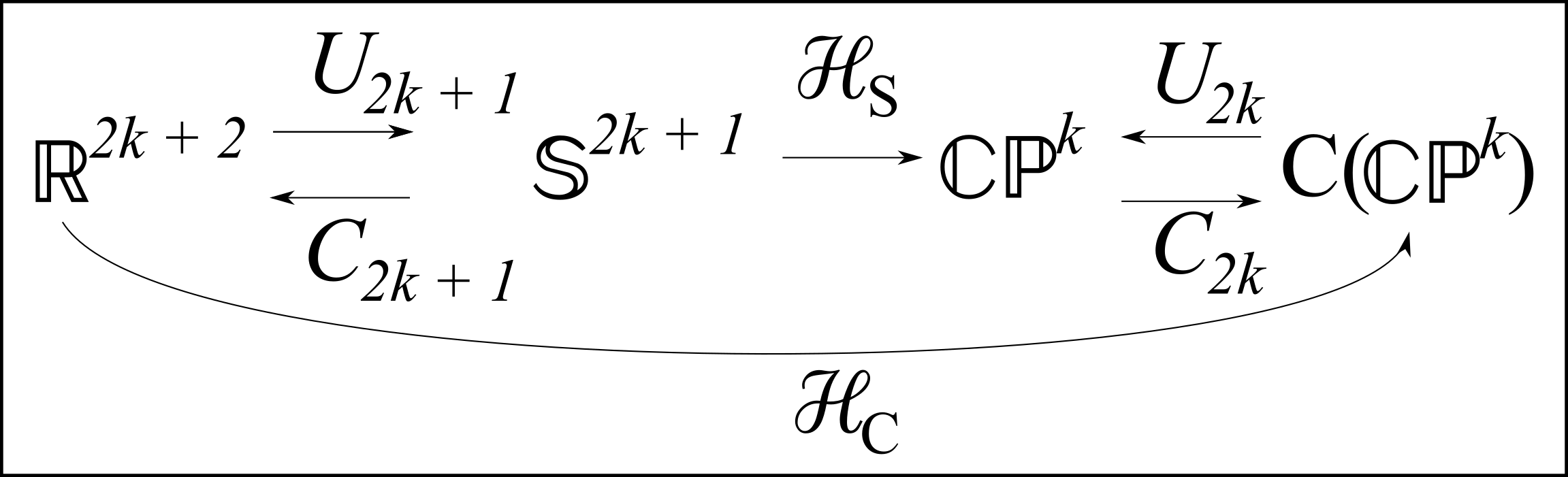}
\caption[Text der im Bilderverzeichnis auftaucht]{        \footnotesize{Hopf map ${\cal H}_{\sS}$ from $\mathbb{S}^{2 \, k + 1}$ to $\mathbb{CP}^k$, 
including mapping the natural ambient $\mathbb{R}^{2 \, k + 2}$ for the $\mathbb{S}^{2 \, k + 1}$ to the cone over  $\mathbb{CP}^k$
 by the Hopf map ${\cal H}_{\sC}$, where the $\mC$ now stands for `cone' rather than `Cartesian'.
$\widetilde{C}_k$ are cone maps from an arbitrary manifold, and $\widetilde{U}_k$ are unit scale maps from the cone back to the manifold being coned over.}  } 
\label{Generalized-Hopf-Map} \end{figure}           }

\ni{\bf Structure 11} See Fig \ref{Generalized-Hopf-Map} for an outline.
Note that Appendix \ref{App-1}'s case is included by virtue of the `accidental relations' 
\be
\mathbb{CP}^1 = \mathbb{S}^2
\ee 
and 
\be
\mC(\mathbb{S}^2) = \mathbb{R}^3 \mbox{ } .
\ee   
\ni{\bf Structure 12} For this generalization, the Hopf quantities are best thought of as $SU(2)$ objects, 
corresponding to the isometry group of $\mathbb{CP}^1 = \mathbb{S}^2$ being 
\be 
\frac{  SU(2)  }{  \mathbb{Z}_2  } \mbox{ }  \left[ \mbox{ } =  SO(3) \mbox{ } \right] \mbox{ } .
\ee
\ni $\mathbb{CP}^k$'s isometry group is moreover \cite{QuadI}
\be
\frac{  SU(k + 1)  }{  \mathbb{Z}_{k + 1}  } \mbox{ } , 
\ee 
and now the analogous Hopf quantities are a set of $(k + 1)^2 - 1 = k(k + 2)$ objects built from $k + 1$ $\mathbb{R}^2$ vectors 
                                            that span the $\mathbb{R}^{2 \, k + 2}$ ambient space of the $\mathbb{S}^{2 \, k + 1}$.  
%
%
See \cite{QuadI} for details of these quantities for $k = 2$.

\section{Shape-theoretic realization of the generalized Hopf map, and generalized Kendall's Theorem}\label{App-4}
%
{            \begin{figure}[!ht]
\centering
\includegraphics[width=0.3\textwidth]{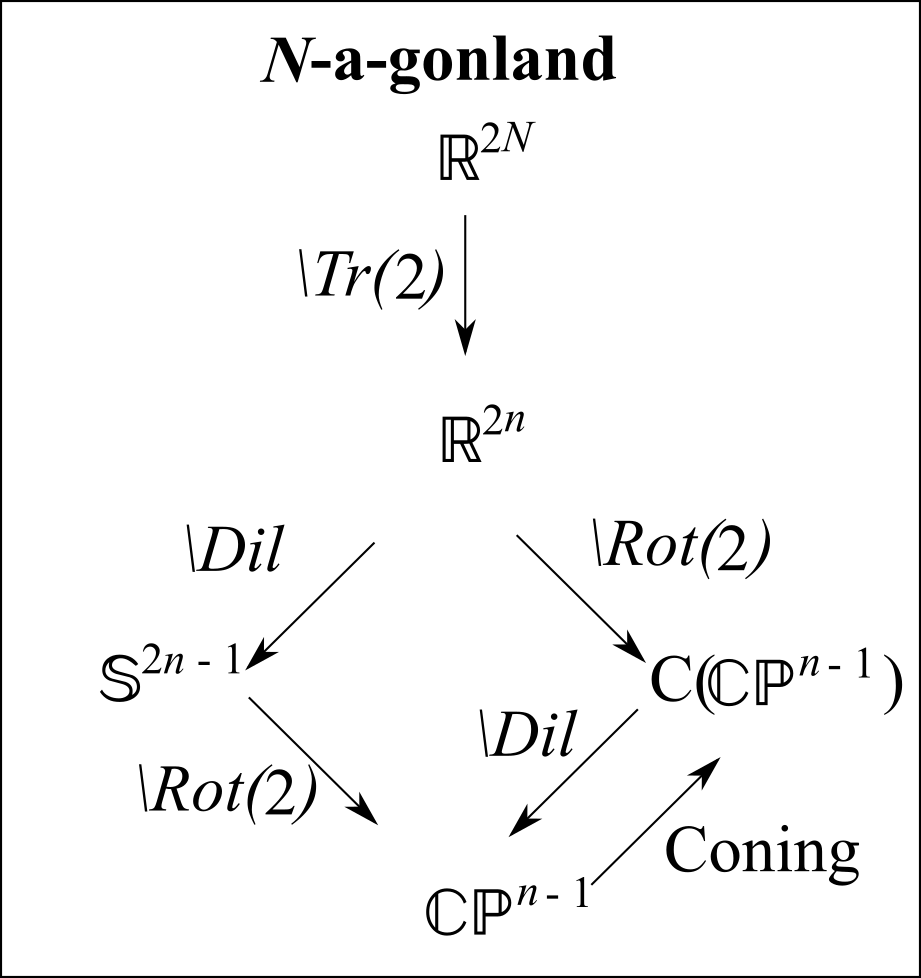}
\caption[Text der im Bilderverzeichnis auftaucht]{        \footnotesize{Configuration spaces for the $N$-a-gon; these realize the generalized Hopf maps 
${\cal H}_{\sS}$ and ${\cal H}_{\sC}$ in direct analogy to how Fig 6 realizes the non-generalized versions of these.}  } 
\label{N-a-gonland} \end{figure}           }

\ni The $k + 1 = n$ $\mathbb{R}^{2}$ vectors are now $\urho_a$, $a = 1$ to $n$.
The above $k = 2$ example's Hopf quantities are furthermore interpreted as shape quantities in the given reference \cite{QuadI}, corresponding to the  $N = 4$ case of quadrilaterals.

\mbox{ }

\ni{\bf Theorem 10 (generalized Kendall's Theorem)} The shape space of $N$-a-gons, in the sense of labelled mirror-image-distinct $N$-point constellations in the plane, 
is topologically $\mathbb{CP}^{N - 2}$ and equipped furthermore at the metric level with the standard Fubini--Study metric.  

\mbox{ }

\ni{\bf Remark 45} The geometrical \cite{Kendall84, Kendall89, Kendall}, mechanics-reduction \cite{FORD, FileR} and Hopf map \cite{III} proofs of this carry through. 
It remains to be seen whether the general $N$-a-gon case admits an area-expression reformulation proof along the lines of the current paper's. 

\end{appendices}

\vspace{10in}


\end{document}